\documentclass[notitlepage,prd,showkeys,twocolumn,preprintnumbers,floatfix,superscriptaddress,nofootinbib]{revtex4-1}
\usepackage{amsmath}
\usepackage{amssymb}
\usepackage{xcolor}
\usepackage{bbm}
\usepackage{braket}
\usepackage{xspace}
\usepackage{mathtools}
\usepackage{graphicx}
\usepackage{enumerate}
\usepackage{hyperref}
\usepackage{verbatim}
\usepackage{tabularx}
\usepackage{multirow}
\usepackage{enumitem}
\usepackage{orcidlink}

\newcommand{\SLJ}[3]{\ensuremath{{\:\!}^{{#1}\!}{#2}_{#3}}}
\newcommand{\vecb}[1]{\boldsymbol{#1}}

\usepackage{etoolbox}
\makeatletter
\patchcmd{\subsubsection}{\itshape}{\itshape\bfseries}{}{}
\makeatother

\newcommand{\2}{\mathbf{2}}

\newcommand{\Cc}{\mathcal{C}}
\newcommand{\Dc}{\mathcal{D}}

\newcommand{\Fc}{\mathcal{F}}
\newcommand{\Gc}{\mathcal{G}}

\newcommand{\Ic}{\mathcal{I}}

\newcommand{\Kc}{\mathcal{K}}
\newcommand{\Lc}{\mathcal{L}}
\newcommand{\Mc}{\mathcal{M}}

\newcommand{\Pc}{\mathcal{P}}
\newcommand{\Qc}{\mathcal{Q}}
\newcommand{\Rc}{\mathcal{R}}
\newcommand{\Tc}{\mathcal{T}}

\newcommand{\wt}[1]{\widetilde{#1}}

\newcommand{\ie}{i.e.\xspace}

\newcommand{\diff}{\textrm{d}}
\newcommand{\df}{\textrm{df}}

\newcolumntype{Y}{>{\centering\arraybackslash}X}
\newcolumntype{s}{>{\centering\arraybackslash\hsize=.2\hsize}X}

\definecolor{wm_green}{HTML}{115740}
\definecolor{wm_gold}{HTML}{B9975B}

\usepackage{mfirstuc}
\newcommand{\addReviewer}[2]{
\expandafter\newcommand\csname #1\endcsname[1]{{\sf \color{#2} {#1}:\,##1}}
\expandafter\newcommand\csname #1cor\endcsname[2]{{\color{#2} {#1}:\,\st{##1}{\sf ##2}}}
\expandafter\newcommand\csname #1color\endcsname{#2}
}

\usepackage{soul,color}
\definecolor{chromeyellow}{rgb}{1.0, 0.65, 0.0}
\definecolor{DodgeBlue}{rgb}{0.118, 0.565,1.000}
\definecolor{asparagus}{rgb}{0.53, 0.66, 0.42}
\definecolor{cadmiumgreen}{rgb}{0.0, 0.42, 0.24}

\definecolor{jlab_red}{RGB}{192,39,45}
\definecolor{jlab_orange}{RGB}{249,102,0}
\definecolor{jlab_blue}{RGB}{47,122,121}
\definecolor{jlab_green}{RGB}{65,125,10}

\addReviewer{mh}{blue}
\addReviewer{rge}{purple}
\addReviewer{ct}{red}
\addReviewer{rb}{jlab_blue}
\addReviewer{aj}{jlab_orange}

\hypersetup{
pdftitle = {rho-pi},
pdfsubject = {Scattering Theory},
pdfkeywords = {Scattering, Partial Waves, QCD, Hadron, Physics, Lattice},
pdfauthor = {Briceno, Hansen, Jackura, Edwards, Thomas},
pdfnewwindow = {true},
colorlinks = {true},
linkcolor = {blue},
citecolor = {blue},
filecolor = {blue},
urlcolor = {blue}
}

\newcommand{\wm}{Department of Physics,
William \& Mary,
Williamsburg, VA 23187, USA}
\newcommand{\ucb}{Department of Physics,
University of California,
Berkeley, CA 94720, USA}
\newcommand{\lbnl}{Nuclear Science Division,
Lawrence Berkeley National Laboratory, Berkeley,
CA 94720, USA}
\newcommand{\jlab}{Thomas Jefferson National Accelerator Facility, 12000 Jefferson Avenue, Newport News, Virginia 23606, USA}
\newcommand{\edinburgh}{Higgs Centre for Theoretical Physics, School of Physics and Astronomy, The University of Edinburgh, Edinburgh EH9 3FD, UK}

\begin{document}

\title{Isotensor $\pi\pi \pi$ scattering with a $\rho$ resonant subsystem from QCD}

\preprint{JLAB-THY-25-4592}


\author{Ra\'ul A. Brice\~no~\orcidlink{0000-0003-1109-1473}}
\email[e-mail: ]{rbriceno@berkeley.edu}
\affiliation{\ucb}
\affiliation{\lbnl}

\author{Maxwell T.~Hansen~\orcidlink{0000-0001-9184-8354}}
\email[e-mail: ]{maxwell.hansen@ed.ac.uk}
\affiliation{\edinburgh}

\author{Andrew~W.~Jackura~\orcidlink{0000-0002-3249-5410}}
\email[e-mail: ]{awjackura@wm.edu}
\affiliation{\wm}

\author{Robert~G.~Edwards~\orcidlink{0000-0002-5667-291X}}
\email[e-mail: ]{edwards@jlab.org}
\affiliation{\jlab}

\author{Christopher~E.~Thomas~\orcidlink{0000-0001-8817-4977}}
\email[e-mail: ]{c.e.thomas@damtp.cam.ac.uk}
\affiliation{DAMTP, University of Cambridge, Wilberforce Road, Cambridge, CB3 0WA, UK}

\collaboration{for the Hadron Spectrum Collaboration}

\begin{abstract}
This work presents a lattice quantum chromodynamics (QCD) determination of $\pi\pi\pi$ scattering amplitudes for the isospin-2 channel with angular momentum and parity $J^{P} = 1^+$. The calculation is performed using unphysically heavy light-quark masses, corresponding to a pion mass of $m_{\pi} \approx 400$~MeV, for which the $\rho$ meson manifests as a narrow resonance. The analysis employs a previously developed formalism that non-perturbatively relates the finite-volume spectra of two- and three-pion systems to their infinite-volume scattering amplitudes. We combine earlier lattice QCD results for $\pi\pi$ systems with isospins 1 and 2 with new determinations of the isospin-2 three-pion finite-volume spectrum, obtained for three cubic, periodic lattice volumes with box length ranging from $\approx 4/m_\pi$ to $6/m_\pi$. These combined data constrain the low-lying partial waves of the infinite-volume three-pion K matrix, from which we solve a set of coupled integral equations to extract the corresponding scattering amplitudes.
\end{abstract}
\date{\today}

\maketitle

\emph{\bf Introduction:}
Although quantum chromodynamics (QCD) is formulated in terms of quarks and gluons, confinement prevents experiments from observing these constituents directly. One instead detects composite hadrons, such as pions, protons and neutrons, and the properties of the underlying theory are inferred from their phenomenology. Precise, first-principles predictions of multi-hadron amplitudes are thus essential to advancing our understanding of nuclear and particle physics.

As an example, low-energy nuclear physics would benefit immensely by applying QCD determinations of multi-nucleon amplitudes to reduce uncertainties in the two-neutron force (currently $\sim10\%$) and even larger uncertainties in three-body forces, which currently hinder the precise determination of nuclear properties~\cite{Gobel:2021pvw}. Various precision tests of the Standard Model, such as neutrino-nucleus scattering~\cite{DUNE:2016hlj}, double-beta decay~\cite{Cirigliano:2022oqy}, and multi-hadron weak decays~\cite{LHCb:2019xmb, LHCb:2022fpg, LHCb:2014mir, LHCb:2019jta, LHCb:2013lcl, Suzuki:1999uc, Wolfenstein:1990ks, Suzuki:2007je, AlvarengaNogueira:2015wpj, Bediaga:2013ela, Garrote:2022uub}, are similarly limited by poor knowledge of multi-hadron dynamics and would thus also clearly profit from reliable ab initio determinations.

Hadron spectroscopy, currently experiencing a remarkably rich period of discoveries, is another area where a deeper understanding of such observables would have a significant impact. A recent example of such a discovery is the doubly-charmed tetraquark candidate, $T_{cc}$, observed to decay as $T_{cc} \to D D^\ast \to D D \pi$~\cite{LHCb:2021vvq}. Understanding the true nature of such states from first principles requires precise three-hadron decay amplitudes.\footnote{For ongoing efforts to study the $T_{cc}$ directly from QCD, see Refs.~\cite{Padmanath:2022cvl,Du:2023hlu,Raposo:2023oru,Raposo:2025dkb,Whyte:2024ihh,Lyu:2023xro,Hansen:2024ffk,Shrimal:2025ues,Dawid:2025wsn}.}
More generally, because most hadrons are unstable resonances that decay to two or more lighter hadrons~\cite{ParticleDataGroup:2024cfk}, such amplitudes are essential in rigorously defining and extracting their properties.

To enhance the potential across all these fields, a concerted lattice QCD effort has been initiated to systematically determine scattering amplitudes from QCD. This is challenged by the fact that numerical lattice QCD is defined in a finite, discretized Euclidean spacetime volume where scattering observables are not directly accessible. Indirect constraints can be obtained from a correspondence between the finite-volume spectrum and the desired amplitudes.
For two-particle systems, this correspondence is well understood
~\cite{%
Luscher:1990ux,%
Rummukainen:1995vs,%
He:2005ey,%
Christ:2005gi,%
Kim:2005gf,%
Lage:2009zv,%
Bernard:2010fp,%
Fu:2011xz,%
Briceno:2012yi,%
Hansen:2012tf,%
Guo:2012hv,%
Hall:2013qba,%
Briceno:2014oea,%
Wu:2014vma,%
Hall:2014uca,%
Wu:2014vma,%
Liu:2015ktc,%
Wu:2016ixr,%
Wu:2017qve,%
Li:2019qvh,%
Meng:2021uhz,%
Yang:2021tvc,%
Li:2021mob,%
Du:2023hlu,%
Abell:2023nex,%
Abell:2023qgj,%
Yu:2023xxf,%
Raposo:2023oru,%
Dawid:2024oey,%
Li:2024zld,%
Stokes:2024haa,%
Raposo:2025dkb,%
Wang:2025hew%
}
and
the determination of low-lying two-hadron scattering amplitudes is now a mature aspect of this program, even in kinematic regimes where multiple two-hadron channels are open~\cite{Dudek:2012xn,Dudek:2014qha,Fahy:2014jxa,Wilson:2014cna,Green:2014dea,Wilson:2015dqa,Bolton:2015psa,Bulava:2015qjz,Junnarkar:2015jyf,Dudek:2016esq,Dudek:2016cru,Bulava:2016mks,Briceno:2016mjc,Moir:2016srx,Doring:2016bdr,Wilson:2016rid,Briceno:2017qmb,Andersen:2017una,Brett:2018jqw,Francis:2018qch,Brett:2018fdb,Andersen:2018mau,Hanlon:2018yfv,Woss:2019hse,Wilson:2019wfr,Bulava:2019hpz,Andersen:2019ktw,Fischer:2020yvw,Woss:2020ayi,Johnson:2020ilc,Green:2021qol,Gayer:2021xzv,Green:2021sxb,Padmanath:2022cvl,Radhakrishnan:2022ubg,Bulava:2022vpq,Green:2022rjj,Bulava:2023wrz,Lyu:2023xro,Draper:2023boj,Rodas:2023gma,Rodas:2023nec,BaryonScatteringBaSc:2023ori,BaryonScatteringBaSc:2023zvt,Wilson:2023anv,Wilson:2023hzu,Bulava:2023uma,Skinner:2023wwb,Bulava:2024bsi,Collins:2024sfi,Yeo:2024chk,Boyle:2024hvv,Whyte:2024ihh,Dudek:2024roh,Boyle:2024grr,Dawid:2024dgy,Vujmilovic:2024snz,Francis:2024fwf,Erben:2025zph,Lang:2025pjq}.

Given this success, increased attention has been directed towards advancing the determination of scattering amplitudes involving three hadrons.
Further motivation comes from the fact that scattering amplitudes involving more than two hadrons generally become more relevant for calculations with close-to-physical-mass light quarks, since three-hadron states generically must be incorporated to rigorously treat center-of-momentum energies above the lowest lying three-hadron threshold.
In the three-particle sector, the correspondence between the finite-volume spectrum and scattering amplitudes is known for a more limited, though ever growing, set of systems~\cite{Polejaeva:2012ut,Hansen:2014eka,Hansen:2015zga,Briceno:2017tce,Guo:2017ism,Hammer:2017uqm,Hammer:2017kms,Mai:2017bge,Doring:2018xxx,Briceno:2018mlh,Klos:2018sen,Briceno:2018aml,Guo:2018ibd,Jackura:2019bmu,Blanton:2019igq,Pang:2019dfe,Romero-Lopez:2019qrt,Blanton:2020gha,Blanton:2020jnm,Romero-Lopez:2020rdq,Muller:2020vtt,Muller:2020wjo,Muller:2021uur,Muller:2022oyw,Jackura:2022xml,Baeza-Ballesteros:2023ljl, Hansen:2020zhy,Pang:2020pkl,Blanton:2020gmf,Blanton:2021mih,Blanton:2021eyf,Draper:2023xvu,Hansen:2025oag,Hansen:2021ofl,Severt:2022jtg,Bubna:2023oxo}.
To date, most systematic constraints of three-hadron scattering dynamics have been for weakly repulsive mesonic systems~\cite{Beane:2007es,Detmold:2008fn,Blanton:2019vdk,Hansen:2020otl,Blanton:2021llb,Dawid:2025doq}.
Three notable exceptions are lattice QCD studies constraining the $\pi\pi\pi$ scattering amplitudes in the $a_1(1260)$~\cite{Mai:2021nul}, $\omega(782)$~\cite{Yan:2024gwp} and $\pi(1300)$~\cite{Yan:2025mdm} channels.
For recent reviews on the general approach, we point the reader to Refs.~\cite{Briceno:2017max, Hansen:2019nir,Mai:2021lwb}.

The calculation presented here advances the goal of reliably predicting multi-hadron amplitudes by presenting the first determination of a $\pi\pi\pi \to \pi\pi\pi$ scattering amplitude with coupled (resonant and non-resonant) two-pion subchannels. In particular, we study the coupled $(\pi\pi)_{I_2=1}\pi$ and $(\pi\pi)_{I_2=2}\pi $ channels, where $I_2$ is the isospin of the two-pion subsystem, within the total isospin $I_3=2$ (isotensor) channel with $J^P = 1^+$, where $J$ and $P$ are total angular momentum and parity. This calculation is performed using a single set of unphysically heavy light-quark masses, corresponding to a pion mass of $m_\pi \approx 400$~MeV~\cite{Dudek:2012gj}, with the strange quark having an approximately physical mass. For these values of the quark masses, the $(\pi\pi)_{I_2=1}$
channel couples strongly to the unstable $\rho$ resonance, which is narrow with a width of $\Gamma\approx10$~MeV~\cite{Dudek:2012xn}.\footnote{The channel considered here has been studied previously for heavier quark masses where the $\rho$ is stable~\cite{Woss:2018irj}.}
In contrast to Refs.~\cite{Mai:2021nul, Yan:2024gwp, Yan:2025mdm}, this work relies on QCD, unitarity, and analyticity constraints, together with a set of parametrizations intended to estimate the systematic uncertainties associated with the unknown true functional forms of the amplitudes. In particular, we do not impose constraints from chiral effective theory or use quark-mass-dependent parametrizations. As we discuss in the outlook, it is not clear whether the former are applicable for this system, with our value of the light-quark mass corresponding to $m_\pi \approx 400 \, \text{MeV}$.

\begin{table}[t]
\renewcommand{\arraystretch}{1.1}
\begin{tabular}{c|ccc}
$(L/a_s)^3 \times (T/a_t)$ &$N_{\rm cfgs}$ & $N_{\textrm{tsrc}}$ & $N_{v}$ \\
\hline
$16^3\times 128$ & 479 & 4 & 64\\
$20^3\times 256$ & 286 & 4 & 128\\
$24^3\times 128$ & 553 & 1 & 160\\
\end{tabular}
\ \ \
\begin{tabular}{c|l}
\multicolumn{2}{c}{}\\
$a_t m_\pi$ & $0.06906(13) \ \ $ \\
$a_t m_K$ & $0.09698(9)$ \\
$a_t m_\omega$ & $0.15541(29)$ \\
\end{tabular}
\renewcommand{\arraystretch}{1.0}
\caption{Details of the lattice ensembles used in this work. \emph{Left:} lattice geometry, number of gauge field configurations ($N_{\rm cfgs}$), number of time sources per configuration ($N_{\textrm{tsrc}}$), and number of distillation vectors ($N_{v}$). \emph{Right:} Relevant stable meson masses in units of the temporal lattice spacing~\cite{Dudek:2012gj,Woss:2019hse,Wilson:2014cna}.\label{TABLE:lattices}}
\end{table}

In the remainder of this work, we describe the key elements of the calculation. We begin with a determination of the finite-volume spectrum, followed by analysis of the spectrum to obtain an infinite-volume scheme-dependent K matrix using the formalism derived and implemented in Refs.~\cite{Hansen:2014eka,Hansen:2020zhy,Alotaibi:2025pxz}.
This is subsequently used to determine physical observables (scattering amplitudes and intensities) by solving integral equations derived in Refs.~\cite{Hansen:2015zga,Hansen:2020zhy, Jackura:2022gib} using advances described in Refs.~\cite{Dawid:2023jrj, Jackura:2023qtp, Briceno:2024ehy, Jackura:2025wbw}. We close by discussing conclusions that can be drawn from the analysis. More details of our methods and their implementation are given in Supplemental Material.

\emph{\bf Spectrum determination:} The spectra of three-hadron systems are substantially more challenging to determine than for two-hadron systems. As with previous two-hadron calculations, we use a state-of-the-art method where a large set of two-point correlation functions involving appropriate interpolating operators is computed in each symmetry channel. Then the spectrum is determined by solving a generalized eigenvalue problem~\cite{Michael:1985ne,Luscher:1990ck,Blossier:2009kd} -- our implementation is described in Ref.~\cite{Dudek:2010wm}.
Here we use operators with $\rho\pi$-like and $\pi\pi\pi$-like structures~\cite{Dudek:2012gj,Woss:2019hse,Hansen:2020otl}, up to 6 or 7 of each depending on the lattice volume.
More details of the spectrum determination and operator constructions are given in Sec.~\ref{app:ops_spec} of the Supplemental Material.
To evaluate the correlation functions, one must compute all possible Wick contractions of the quark fields, including those involving quark-antiquark annihilation. We do this efficiently using the distillation method~\cite{Peardon:2009gh}.

The calculations are performed on the anisotropic set-up detailed in Ref.~\cite{HadronSpectrum:2008xlg}, with three cubic lattices differing only in the spatial extents $L$ corresponding to $m_\pi L \approx 3.8, 4.8, 5.7$ and temporal extents $T$, given by $m_\pi T\approx 8.8, 17.8, 8.8$, respectively.
The temporal lattice spacing is
$a_t = 0.0348(3)$ fm, where the omega baryon mass~\cite{Edwards:2012fx} is used to set the scale, and the anisotropy (ratio of spatial and temporal lattice spacings) is
$\xi = a_s/a_t = 3.444(6)$~\cite{Dudek:2012gj}.
Other details of the lattices, including masses of relevant mesons, are given in Table~\ref{TABLE:lattices}.
From these, we note that there are two additional relevant thresholds between the three- and five-pion thresholds: $E^{\sf thr}(K\bar{K}\pi)/m_\pi = 3.81$ and $E^{\sf thr}( \omega \pi \pi)/m_\pi = 4.25$. For the lower of these two, we have additionally evaluated correlators with $K\bar{K}\pi$ operators to confirm that the effect of these is negligible (more quantitatively the shift in central values is much less than the statistical uncertainty) in the energy region we are considering.
This is consistent with the findings of Ref.~\cite{Wilson:2015dqa}.
Here it is also relevant to note that the lowest lying non-interacting energies in the neglected channels are significantly above the indicated thresholds, as two of the hadrons must carry back-to-back momentum to achieve the quantum numbers of interest.

A technically distinct but related question is the effect of neglecting these channels in the implementation of the three-particle formalism used to extract the amplitudes. If the coupling between the channel of interest and the neglected channels is sufficiently weak, then the associated effects will be subdominant to the statistical uncertainties reported for our final results. The observation that adding the corresponding operators shifts the finite-volume energies by amounts well below their statistical uncertainties makes it plausible that this hierarchy of uncertainties is preserved in the final amplitudes. In addition, the fact that the fits applied in the next section successfully describe the data using a formalism that neglects the additional channels gives further support to this conclusion. Still, future work is needed to address this point more directly.

The reduced symmetry of the finite-volume discretized system, as compared to an infinite-volume continuum, means that angular momentum is no longer a good quantum number and states are instead labeled by irreps, $\Lambda$, of the cubic group or a subgroup. To avoid complications in the subsequent analysis, we restrict ourselves to systems with total momentum zero in this first study. This ensures that parity, $P$, is a good quantum number, reducing the number of partial waves that can mix. For $J^P=1^+$, the relevant lattice irrep is then $\Lambda^P=T_1^+$, which also includes contributions from $J^P=3^+$ and higher $J$. At low energies, we expect this system to be saturated by the $J=1$ $\pi\pi\pi$ system since the other contributions are kinematically suppressed at low energies by additional barrier factors~\cite{Briceno:2024ehy}.

\begin{figure}[t]
\begin{center}
\includegraphics[width=.485\textwidth]{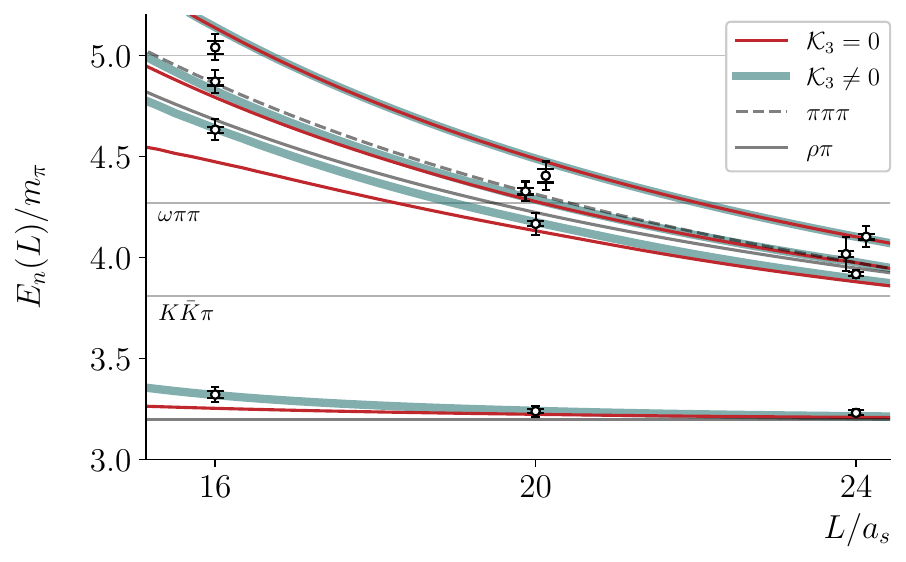}
\caption{Energies in the isospin-2 three-pion channel plotted as a function of the box size $L$. The black points represent lattice QCD-computed energies; inner error bars indicate statistical uncertainties, while outer error bars also include systematic uncertainties as discussed in Sec.~\ref{app:fits} of the Supplemental Material. The red curves are predictions of the finite-volume quantization condition with a vanishing three-body K-matrix \( \mathcal{K}_{3} = 0 \), using two-pion K-matrices from a previous analysis of the two-pion spectrum, while the blue curves show results incorporating a best-fit \( \mathcal{K}_{3} \) as detailed in the text. The horizontal grey lines indicated neglected three-hadron thresholds while the energy-dependent grey lines correspond to the hypothetical spectrum of non-interacting hadrons, as explained in the text. The first excited grey curve has a two-fold multiplicity. \label{fig:spec}}
\end{center}
\end{figure}

In Fig.~\ref{fig:spec}, we show the four lowest energy levels obtained in the $T_1^+$ irrep for each of the three volumes. In addition to presenting the lattice QCD spectra, we show various volume-dependent energy curves predicted by our formalism as described in the caption.

We note that the level counting is consistent with the expectations from a theory of non-interacting pions and rho mesons. The ground state corresponds to a rho and a pion, each at rest, while the first two excited states correspond to the same configuration with one unit of back-to-back momentum. The third excited state corresponds to three pions, two of which carry one unit of back-to-back momentum. We represent this in Fig.~\ref{fig:spec} by plotting in grey the appropriate functions, given by summing
individual particle energies for the relevant momenta. In addition to the pion mass, here we require a choice for our hypothetical non-interacting $\rho$. We set its mass to $ 2.184 m_\pi$, which is the central value of the Breit-Wigner mass, determined from an analysis of the two-particle spectrum of this system in Ref.~\cite{Dudek:2012xn}.
We now go beyond this simple counting to a detailed analysis of the interacting theory implied by this spectrum.

\emph{ \bf Spectrum analysis:}
All formalisms available to date that connect finite-volume spectra to infinite-volume three-hadron scattering, first relate the spectrum to intermediate functions that describe the short-distance dynamics, denoted in the formalism that we use here as K matrices. These are known to have exponentially suppressed volume effects that we neglect, and are real and smooth in the kinematic region considered.

The relation in the $I_3=2$ channel can be written as a determinant condition over all degrees of freedom that are not fixed by the system's energy~\cite{Hansen:2020zhy}
\begin{align}
\det_{I_2,\boldsymbol{k},S,m}\left[ \textbf F_3^{-1}(E_n, L \, \vert \, \boldsymbol \eta_2) + \textbf K_3(E_n \, \vert \, \boldsymbol \eta_3) \right] = 0 \, .
\label{eq:QC}
\end{align}

Before giving some details of this equation, we explain the essential idea. For a given $\textbf K_3$ and for given two-pion dynamics (encoded by $\boldsymbol \eta_2$, a vector of two-pion scattering parameters), solutions of Eq.~\eqref{eq:QC} predict the finite-volume spectrum $\{E_n\}$. We fix $\boldsymbol \eta_2$ from previous calculations \cite{Dudek:2012gj, Dudek:2012xn} and parameterize $\textbf K_3$ with parameters $\boldsymbol \eta_3$ to predict $\{E_n(L \, \vert \, \boldsymbol \eta_2, \boldsymbol \eta_3)\}$. We then form a chi-squared with these predictions and the lattice-QCD-determined energies and vary $\boldsymbol \eta_3$ to minimize, thereby determining the best description.

Both matrices in Eq.~\eqref{eq:QC} are indexed by $I_2 \boldsymbol kS m$ for the initial state and a second primed set for the final state. $ \textbf K_3 $ is the three-body K matrix, and $ \textbf F_{3}$ is a finite-volume matrix that also depends on two-pion interactions.
Here $S$, $m$, and $I_2$ denote, respectively, the angular momentum, $z$-component of angular momentum, and isospin of one of the pairs in the initial state, while $\boldsymbol k$ denotes the momentum of the third particle. We refer to the two pions labeled by $S$, $m$ and $I_2$ as the dipion, and the third as the spectator.

In Sec.~\ref{app:spec_anal} of the Supplemental Material, we provide a detailed description of the definitions of $\textbf K_3$ and $\textbf F_3$ for this channel.
As explained in Refs.~\cite{Hansen:2015zga,Hansen:2020zhy}, $\textbf K_3$ can be converted to a function of three incoming and three outgoing momenta by contracting the $S,m$ indices with spherical harmonics. An important freedom in the definition of $\textbf K_3$ determines whether or not the resulting function is invariant under simultaneous exhange of pion momentum and flavor labels.
This choice is referred to as the symmetric or asymmetric formalism~\cite{Jackura:2022gib}, though we emphasize it only enters the intermediate scheme dependent quantity and either must lead to a final amplitude that is consistent within uncertainties. Here we use the asymmetric formalism. As was shown in Ref.~\cite{Briceno:2024ehy}, this naturally accommodates an arbitrarily large number of parametrizations and thus provides a better handle on systematic errors.

\begin{figure}[t]
\begin{center}
\includegraphics[width=.45\textwidth]{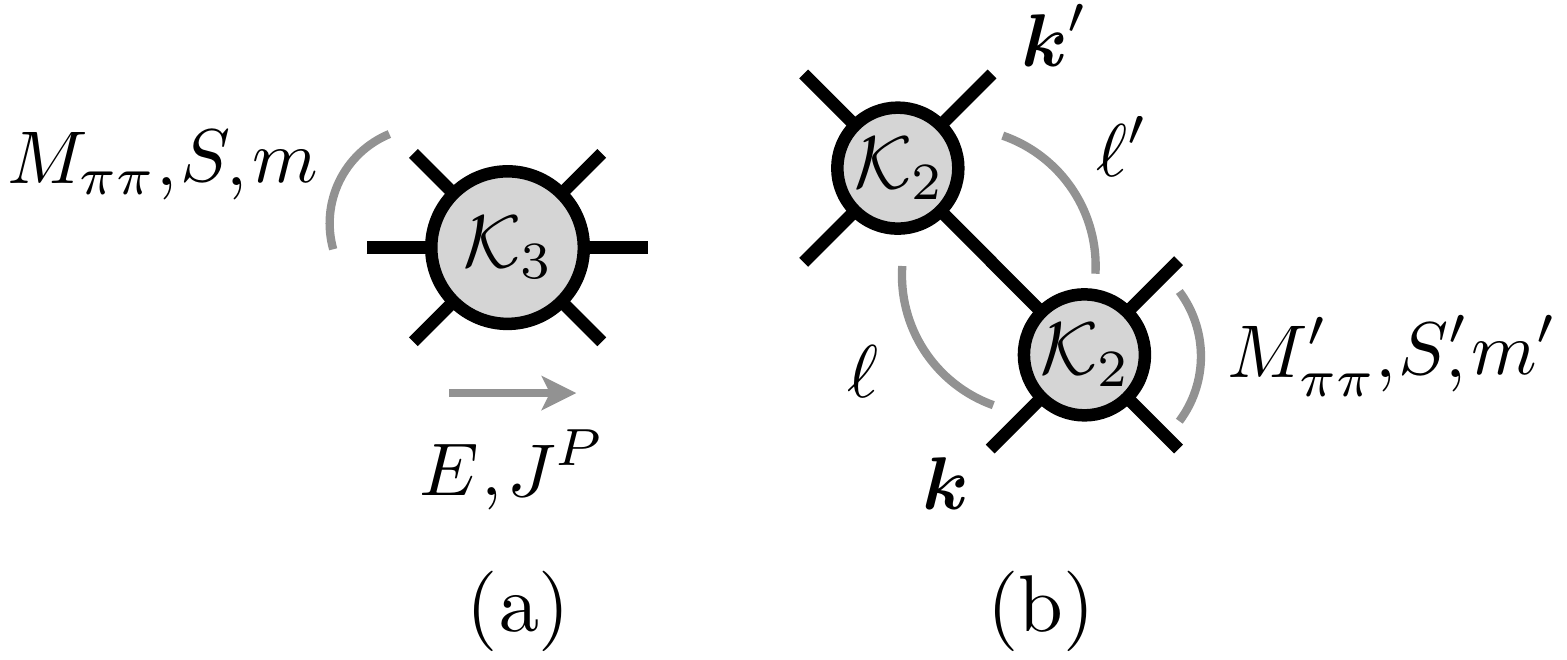}
\caption{Representative diagrams contained in the non-perturbative scattering amplitude, where the black lines represent pions. We show the two leading contributions: (a) the three-body K matrix constrained in this work, and (b) the one-pion exchange diagram responsible for additional kinematic singularities in the amplitude. Kinematic quantities and angular momenta are defined in the text.\label{fig:diagrams}
}
\end{center}
\end{figure}

As briefly indicated above, a key input in the definition of $\textbf F_3$ are the two-body K matrices for the two dipion channels ($I_2=1,2$), which have been previously constrained in Refs.~\cite{Dudek:2012gj, Dudek:2012xn}. For the kinematic region considered, these K matrices, denoted $\Kc_2$ are dominated by the $S=1,0$ partial waves, respectively.\footnote{We keep the scattering parameters of these channels fixed to their mean values and do not propagate their uncertainties which are subdominant to the uncertainties from three-pion finite-volume energies.}

\begin{figure*}[t]
\begin{center}
\includegraphics[width=.85\textwidth]{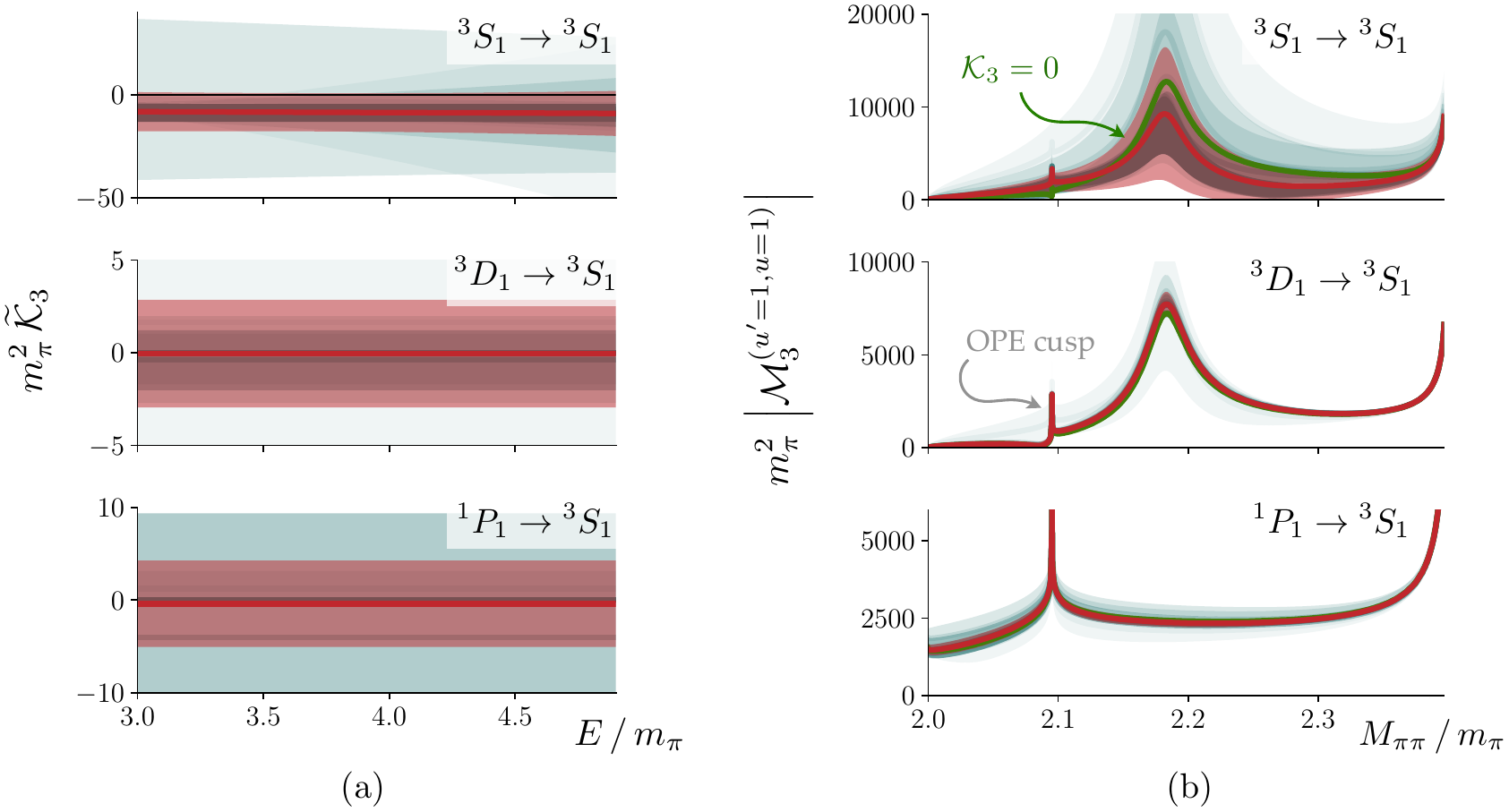}
\caption{
(a) The light blue bands show fit results for a given parametrization of the reduced K matrix as a function of total energy $E$ in the channels $\{{}^3S_1,{}^3D_1,{}^1P_1\} \to {}^3S_1$. Bands represent $1\sigma$ uncertainty of a given fit, and the opacity of the band is directly proportional to the likelihood of the fit, as given by the AIC described in the text, with the darker bands being preferred. The red is the weighted average of the different fits, with the band including the propagated uncertainty. (b) Shown are the $\{{}^3S_1,{}^3D_1,{}^1P_1\} \to {}^3S_1$ unsymmetrized amplitudes as a function of the incoming dipion invariant mass for a fixed $E=3.4 m_\pi$ and outgoing state defined by $M_{\pi\pi}' = 2.1 m_\pi$. The red band includes the error propagation of the K matrices shown in (a). The green curve corresponds to amplitudes for $\Kc_3=0$. The cusp near $2.1m_\pi$ is due to physical OPE.\label{fig:Kmat_amp}
}
\end{center}
\end{figure*}

Given $\textbf F_3$ and $\textbf K_3$, we subduce to the $T_1^+$ irreducible representation of the cubic group, following Refs.~\cite{Blanton:2019igq,Alotaibi:2025pxz} for $\textbf F_3$ and extending the procedure here for the asymmetric $\textbf K_3$. The $T_1^+$ spectrum is expected to be well described by the $J=1$ three-pion system at low energies, and we restrict to channels with $J^P=1^+$ and
$\ell\leq 2$, namely \SLJ{3}{S}{1}, \SLJ{3}{D}{1}, and \SLJ{1}{P}{1}, where we are using the spectroscopic notation \SLJ{2S+1}{\ell}{J} with $\ell$ being the angular momentum between the dipion and spectator pion. See Fig.~\ref{fig:diagrams} for a pictorial representation of the various angular momenta. For these channels, the dipion will have $I_2=1$ in the \SLJ{3}{S}{1} and \SLJ{3}{D}{1} and $I_2=2$ for the \SLJ{1}{P}{1}.

The strategy for parametrizing the three-pion K matrix in this basis follows Ref.~\cite{Briceno:2024ehy}. We consider 10 parametrizations, fit to the lattice QCD spectra of Fig.~\ref{fig:spec}, and perform a model average using the Akaike information criterion (AIC)~\cite{Jay:2020jkz,Pefkou:2021fni}.
Figure~\ref{fig:Kmat_amp}(a) shows the results for the so-called reduced K matrix, $\widetilde{\Kc}_3$,
as a function of the total energy $E$,
for all initial state channels going to the final state ${}^3S_1$.
The reduced K matrix, introduced in Ref.~\cite{Briceno:2024ehy}, is a matrix in channel space and is defined by the decomposition $\mathbf{K}_3 \equiv \mathbf{h}\cdot \wt{\Kc}_3\cdot \mathbf{h}^{\top}$ where $\mathbf{h}$ are known matrices which convert from channel to $I_2\vecb{k}Sm$ space.
We find that after propagating systematic errors, all of the components of the K matrix are consistent with zero.
The best fit for the K matrix is the first listed in Table~\ref{TABLE:fits} of the Supplemental Material. It is reached by setting the $\SLJ{3}{S}{1} \to \SLJ{3}{S}{1}$ component to a floating constant, while all other components of $\widetilde{\Kc}_3$ are set to 0.

\emph{\bf Amplitude analysis:} Having constrained the K matrix, we focus on the determination of the physical observables, namely scattering amplitudes and intensities. The K matrix is generally a scheme-dependent quantity that can be mapped to physical amplitudes via integral equations.\footnote{The scheme-dependence emerges due to a necessary choice of cut-off function for the integral equations and a parametrization of the two-body K matrix.} These were first derived in Ref.~\cite{Hansen:2015zga} for three scalar particles. Using the work of Ref.~\cite{Hansen:2020zhy}, one can immediately write the case for generic $\pi\pi\pi$ systems, and Refs.~\cite{Jackura:2023qtp, Briceno:2024ehy} show how these can be projected to arbitrary angular momentum and parity, yielding the specific form we use here.

\begin{figure*}[t]
\begin{center}
\includegraphics[width=\textwidth]{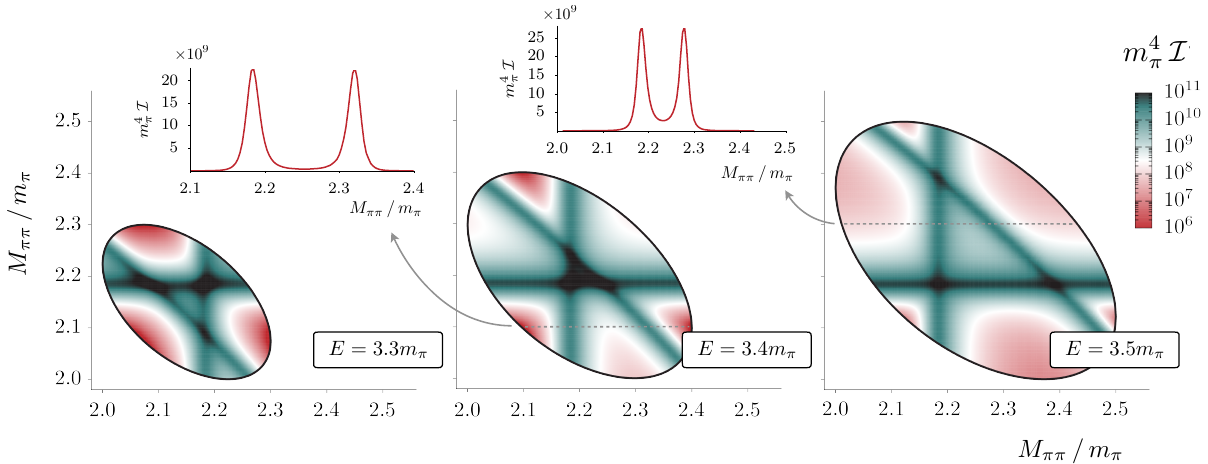}
\caption{\label{fig:Dalitz}Dalitz plots of the intensity distributions in Eq~\eqref{eq:int} for the $J^P=1^+$ amplitude for $E/m_\pi = \{3.3,3.4,3.5\}$.
The axes represent the invariant masses corresponding to two different choices for the dipion within a given state.
Initial and final state kinematics are fixed to be the same. Amplitudes are determined from the best-fitted values of the K matrix described in the text. The insets show slices of the distribution for fixed $(M_{\pi\pi},E) = (2.1m_\pi,3.4m_\pi)$ and $(2.3m_\pi,3.5m_\pi)$. }
\end{center}
\end{figure*}

This framework first yields unsymmetrized $J^P = 1^+$ amplitudes, denoted by $\Mc_3^{(a, b)}$. As for $\widetilde {\mathcal K}_3$ this refers to the fact that, when this intermediate quantity is written as a function of pion momenta and flavor labels, it is not invariant under their simultaneous interchange (as any physical quantity must be). This issue is corrected in a final symmetrization step described below.

Each amplitude is a function of total three-pion energy $E$ and initial and final dipion invariant masses $M_{\pi\pi}$ and $M_{\pi\pi}'$.\footnote{The dipion invariant mass is related to the third pion momentum via $M_{\pi\pi}^2 = E^2 + m_\pi^2 - 2E\sqrt{m_\pi^2 + \vecb{k}^2}$, with similar expression for $M_{\pi\pi}'$ in terms of $\vecb{k}'$.} The indices $a,b \in \{1,2,3\}$ indicate one of the three pions. Schematically, the integral equations depend on two inputs, the two- and three-body K matrices:
\begin{align}
\label{eq:int_eq_schematic}
\Mc_3^{(a,b)} = \Mc_3^{(a,b)}[\Kc_2,\Kc_3] \,.
\end{align}
The driving terms of these integral equations are $\Kc_3$ and the one-pion exchange (OPE) diagram, shown schematically in Fig.~\ref{fig:diagrams}. As with the spectrum, we use the previously constrained elastic $\pi\pi$ $\Kc_2$ for isospin-1 and 2 from Refs.~\cite{Dudek:2012gj, Dudek:2012xn} as an input.
The partial wave $\Kc_3$ used in the integral equations is related to $\mathbf{K}_3$ by usual partial wave projection~\cite{Briceno:2024ehy}, see Sec.~\ref{app:K3_constrain} of the Supplemental Material for further discussion and its relation to $\wt{\Kc}_3$.

We solve the system of integral equations numerically for the different components of $\Mc_3^{(a,b)}$ at a fixed $E$. Results are shown in Fig.~\ref{fig:Kmat_amp}(b) for the $\{{}^3S_1,{}^3D_1,{}^1P_1\} \to {}^3S_1$ amplitudes as a function of the initial dipion invariant mass, $M_{\pi\pi}$, for fixed $E=3.4 m_\pi$ and a final state $M_{\pi\pi}' = 2.1m_\pi$. The $1\sigma$ band is generated by propagating the weighted uncertainties from the 10 fits. For comparison, we also show the amplitudes in the limit that $\Kc_3 = 0$. As can be seen, depending on the channel and kinematics, for a particular fit, there might be statistically significant deviations between the $\Kc_3 = 0$ solution and the full amplitude.\footnote{This is also apparent in Fig. \ref{fig:spec} where the lattice QCD-computed energies are inconsistent with the energies from a parametrization where $\Kc_3 = 0$.} However, when we propagate the error associated with the choice of the parametrization, we find that the $\Kc_3 = 0$ solution and the full amplitude are consistent within uncertainties. 
Notice the ${}^3S_1$ and ${}^3D_1$ channels are dominated by the presence of the $\rho$ resonance, with its peak at $M_{\pi\pi} \approx 2.2m_\pi$, while the ${}^1P_1$ channel is dominated by the logarithmic singularity at $M_{\pi\pi} \approx 2.1m_\pi$ associated with one-pion exchange (OPE).

Finally, we form the fully symmetric amplitude by summing over all spectators choices from the results of Eq.~\eqref{eq:int_eq_schematic} using the procedure presented in Ref.~\cite{Jackura:2025wbw},
\begin{align}
\Mc_{3,I_2'\lambda',I_2\lambda}= \sum_{a,b}{\Rc}_{I_2'\lambda'}^{(a)} \cdot {\Mc}_{3}^{(a,b)} \cdot {\Rc}_{I_2 \lambda}^{(b)\,\top} \, .
\end{align}
Here $\lambda,\lambda'$ are helicity quantum numbers formed by linear combinations of the participating channels, and $\Rc_{I_2 \lambda}^{(a)}$ are recoupling matrices in isospin and partial waves from the $a$th spectator to the desired quantum numbers. Their explicit expression is given in Ref.~\cite{Jackura:2025wbw}.

Reference~\cite{Jackura:2025wbw} also provided a suitable definition for intensities for three-body systems with arbitrary angular momentum and isospin,
\begin{align}
\mathcal{I} =\sum_{I_2',I_2}\sum_{\lambda,\lambda'} \Big\lvert \Mc_{3,I_2'\lambda',I_2\lambda} \,
\Big\rvert^2 .
\label{eq:int}
\end{align}
In Fig.~\ref{fig:Dalitz} we show Dalitz plots of the intensities for $E/m_\pi = 3.3, 3.4, 3.5$. We choose the initial state kinematic configuration to mirror those of the final state. The most striking features are dense bands due to the $\rho$ resonance at $M_{\pi\pi} \approx 2.2 m_\pi$. Because the amplitude is symmetrized, one sees reflections of the $\rho$ bands which constructively interfere in the kinematic region where two bands overlap. Note that the logarithmic singularity that is visible in Fig.~\ref{fig:Kmat_amp} is unresolvable in the intensities. This is due in part to the fact that the intensity is dominated by the ${}^3S_1$ amplitude, which is itself dominated by the $\rho$ peak, and for the matrix sizes for which we were able to solve the integral equations, the $S=1$ softened OPE singularity is too narrow to be resolved~\cite{Jackura:2023qtp}. See Sec.~\ref{app:amp_recon} in the Supplemental Material for a detailed discussion of this point and other aspects of the amplitude analysis.

\emph{\bf Outlook:} In this work, we have presented the first determination of the full kinematic dependence of a three-hadron scattering amplitude for a coupled-channel system with a resonant subsystem. Achieving this result required the development of a general, flexible formal and computational machinery, which can now be applied to the full set of $\pi \pi \pi$ channels of immediate physical interest, including those with three-pion resonances.

One promising future direction of this program is to map out the pion-mass dependence of such scattering amplitudes from very high pion masses down to the physical point. This will provide new insight into the analytic properties and potential universality across different channels. Together with the previously published $\rho\pi$ scattering amplitudes at $m_\pi \sim 700$~MeV~\cite{Woss:2018irj}, this work at $m_\pi \sim 400$~MeV provides a second value with which to begin such an investigation. As one pushes to lighter values of $m_\pi$, chiral perturbation theory ($\chi$PT) can be compared to the data, at least for near-threshold kinematics, to more fully understand this dependence as well as to cross-check the lattice calculation. A relevant formal contribution in this direction was made by Ref.~\cite{Baeza-Ballesteros:2024mii}, which derived the leading-order $\chi$PT expressions for the symmetric $\Kc_3$ at threshold across all $\pi\pi\pi$ isospins. Given the presence of the narrow, near-threshold $\rho$ in the present work, we do not expect those results to directly apply here. In the context of two-pion scattering, unitarized extensions of $\chi$PT~\cite{Dobado:1996ps,Oller:1998hw,GomezNicola:2001as,Pelaez:2006nj} have proven valuable for studying the $\rho$ resonance in lattice QCD~\cite{Bolton:2015psa, Niehus:2021iin}, and similar methods could be explored for three-pion systems in future work. We also note that $\chi$PT has been successfully applied for maximal-isospin three-hadron systems in Refs.~\cite{Dawid:2025doq,Dawid:2025zxc}.

An important conclusion of this calculation is that varying the parametrizations of the three-body K matrix gives an essential handle on the systematic uncertainty that arises from not knowing the full analytic form. In other words, using an insufficient set of parametrizations risks underestimating uncertainties.~\footnote{For example, using fit 1 in Table~\ref{TABLE:fits}, one finds a non-zero value of $\Kc_3$. After varying the choice of the parametrization, $\Kc_3$ becomes consistent with $0$.} We find that the asymmetric formalism makes the allowed forms of $\mathcal K_3$ more transparent, and thereby offers the flexibility needed for robust uncertainty quantification.

The present calculation applies the formalism first derived in Ref.~\cite{Romero-Lopez:2019qrt}, initiated in Refs.~\cite{Hansen:2014eka,Hansen:2015zga}, and later generalized to $2 \to 3$ processes~\cite{Briceno:2017tce} as well as systems with non-degenerate particles and spin~\cite{Blanton:2020gmf,Blanton:2021mih,Blanton:2021eyf,Draper:2023xvu,Hansen:2024ffk,Hansen:2025oag}. Together with recent advances connecting K matrices to amplitudes~\cite{Dawid:2023jrj, Jackura:2023qtp, Briceno:2024ehy, Jackura:2025wbw}, these developments will enable an ambitious program of applications, including $\omega(782)\to 3\pi$, $a_1(1260)\to\rho\pi\to 3\pi$, the Roper resonance $N(1440)\to N\pi\pi$, and fully coupled three-hadron systems such as $\Sigma\pi\pi$ and $K\bar K N$ in the hyperon sector. Such calculations will provide novel model-independent links between QCD's quark--gluon dynamics and complex low-energy phenomena, laying the groundwork for a systematic, first-principles knowledge of the multi-hadron landscape.

\section*{Acknowledgments}

We thank our colleagues within the Hadron Spectrum Collaboration (www.hadspec.org) for useful discussions, and would particularly like to thank J.~Dudek, F.~Ortega-Gamma, and D.~Wilson, for helpful comments on a previous version of this manuscript. We thank N.~Chambers, C.~S.~R.~Costa, S.~Dawid, D.~Pefkou, F.~Romero-L\'opez, S.~R.~Sharpe, J.~Sitison, and R.~Urek for useful conversations related to the analysis presented.
This work contributes to the goals of the U.S. Department of Energy \emph{ExoHad} Topical Collaboration (www.exohad.org), Contract No. DE-SC0023598, and AWJ acknowledges support from the contract.
AWJ was also supported in part by the William \& Mary Faculty Research Grant.
RAB was partly supported by the U.S. Department of Energy, Office of Science, Office of Nuclear Physics under Award No. DE-SC0025665 and No. DE-AC02-05CH11231.
MTH is supported in part by U.K. Science and Technology Facilities Council (STFC) grants ST/T000600/1 and ST/X000494/1 and additionally by UKRI Future Leader Fellowship MR/T019956/1.
RGE acknowledges support from the U.S. Department of Energy contract DE-AC05-06OR23177, under which Jefferson Science Associates, LLC, manages and operates Jefferson Lab.
CET acknowledges support from STFC [grant number ST/X000664/1].
The authors acknowledge support from the U.S. Department of Energy, Office of Science, Office of Advanced Scientific Computing Research and Office of Nuclear Physics, Scientific Discovery through Advanced Computing (SciDAC) program.
Also acknowledged is support from the Exascale Computing Project (17-SC-20-SC), a collaborative effort of the U.S. Department of Energy Office of Science and the National Nuclear Security Administration.

This work used clusters at Jefferson Laboratory under the USQCD Initiative and the LQCD ARRA project.
This work also used the DiRAC Data Intensive service (CSD3) at the University of Cambridge, managed by the University of Cambridge University Information Services on behalf of the STFC DiRAC HPC Facility (www.dirac.ac.uk). The DiRAC component of CSD3 at Cambridge was funded by BEIS, UKRI and STFC capital funding and STFC operations grants. DiRAC is part of the UKRI Digital Research Infrastructure.

Also used was an award of computer time provided by the U.S.\ Department of Energy INCITE program and supported in part under an ALCC award, and resources at: the Oak Ridge Leadership Computing Facility, which is a DOE Office of Science User Facility supported under Contract DE-AC05-00OR22725; the National Energy Research Scientific Computing Center (NERSC), a U.S.\ Department of Energy Office of Science User Facility located at Lawrence Berkeley National Laboratory, operated under Contract No. DE-AC02-05CH11231; the Texas Advanced Computing Center (TACC) at The University of Texas at Austin; the Extreme Science and Engineering Discovery Environment (XSEDE), which is supported by National Science Foundation Grant No. ACI-1548562; and part of the Blue Waters sustained-petascale computing project, which is supported by the National Science Foundation (awards OCI-0725070 and ACI-1238993) and the state of Illinois. Blue Waters is a joint effort of the University of Illinois at Urbana-Champaign and its National Center for Supercomputing Applications.

The software codes
{\tt Chroma}~\cite{Edwards:2004sx}, {\tt QUDA}~\cite{Clark:2009wm,Babich:2010mu}, {\tt QUDA-MG}~\cite{Clark:SC2016}, {\tt QPhiX}~\cite{ISC13Phi},
{\tt MG\_PROTO}~\cite{MGProtoDownload}, {\tt QOPQDP}~\cite{Osborn:2010mb,Babich:2010qb} and {\tt Redstar}~\cite{Chen:2023zyy} were used.

\bibliography{bibi.bib}

\onecolumngrid
\newpage

\appendix

\renewcommand{\appendixname}{}
\renewcommand{\thesection}{\Alph{section}}

\section*{Supplemental Material}

\section{Operator construction and Lattice QCD spectrum}
\label{app:ops_spec}

This work concerns three-pion scattering in the isotensor, axial-vector channel ($I_3=2$, $J^P = 1^+$). For finite-volume systems with zero total momentum in the finite-volume frame, $J^P = 1^+$ appears in the $\Lambda^P = T_1^+$ irreducible representation (irrep) of the octahedral group ($\mathrm{O}_h$). Thus, to determine the finite-volume spectrum of interest
on each lattice QCD gauge ensemble,
we compute correlation functions involving interpolating operators with the relevant quantum numbers ($I_3=2$, negative $G$-parity, total momentum $\boldsymbol P = \boldsymbol 0$, and $\Lambda^P = T_1^+$) and a structure resembling $\rho\pi$ or $\pi\pi\pi$ constructed using the general approach outlined in Ref.~\cite{Dudek:2012gj}.\footnote{A single-meson-like fermion-bilinear operator cannot be constructed with this isospin.}

Following Refs.~\cite{Woss:2019hse,Hansen:2020otl},
the $\pi\pi\pi$ operators are constructed by combining an isovector or isotensor $\pi \pi$ operator ($I_{2} = 1$ or $2$) with an additional single-pion operator. The $\pi\pi$ operators are defined as
\begin{equation}
(\boldsymbol \pi \boldsymbol \pi)^{[\vecb{k}_1,\vecb{k}_2]\dagger}_{I_{2}, \, \Lambda_{12}, \, \mu_{12}}(\vecb{k}_{12}) =
\sum\limits_{\substack{\vecb{k}_1, \, \vecb{k}_2 \\ \vecb{k}_1+\vecb{k}_2 = \vecb{k}_{12}}}
\mathcal{C}(\vecb{k}_{12}, \Lambda_{12}, \mu_{12}; \vecb{k}_1, \Lambda_1; \vecb{k}_2, \Lambda_2) \;
\boldsymbol \pi^\dagger_{\Lambda_1}(\vecb{k}_1) \cdot \boldsymbol C_{\sf iso}(I_2, 1, 1) \cdot \boldsymbol \pi^\dagger_{\Lambda_2}(\vecb{k}_2) \, ,
\label{equ:2piops}
\end{equation}
where the sum is over all momenta related to $\vecb{k}_1$ and $\vecb{k}_2$ by allowed lattice rotations, and $\mathcal{C}$ is an appropriate generalized Clebsch-Gordan coefficient for $\Lambda_1 \otimes \Lambda_2 \to \Lambda_{12}$.
We represent the projection to definite two-pion isospin compactly by introducing $\boldsymbol C_{\sf iso}(I_2, 1, 1)$ as a rank-three tensor in the isospin space, populated with the Clebsch-Gordan coefficients for $1 \otimes 1 \to I_2$. On the right-hand side of the equation, two of the indices are contracted with
the single-pion operators, each understood as three-component vectors in isospin. The third implicit index of $\boldsymbol C_{\sf iso}$ gives the isospin row index for the two-pion operator the left-hand side.
The single-pion operators are themselves an optimized linear combination, to interpolate the groundstate with momentum $\vecb{k}_i$ in a given irrep\footnote{These are all one-dimensional finite-volume irreps and thus we omit the irrep row index.}
$\Lambda_i$ using a basis of fermion-bilinear operators featuring various Dirac $\gamma$ matrices and gauge-covariant derivatives -- see Ref.~\cite{Dudek:2012gj} for details. In this work the basis of fermion-bilinear operators used for a single-pion operator has up to three derivatives for zero momentum and up to one derivative for non-zero momentum, except we use up to two derivatives for $1 \leq |L \vecb{k}_i/(2 \pi)|^2 \leq 4$ on the $24^3$ volume.

As in Eq.~(4) of Ref.~\cite{Woss:2019hse}, the $\pi\pi\pi$ operators are then defined as
\begin{equation}
(\boldsymbol \pi \boldsymbol \pi \boldsymbol\pi)^{[I_{2}, \Lambda_{12};\vecb{k}_{12}[\vecb{k}_1,\vecb{k}_2],\vecb{k}_3]\dagger}_{\Lambda, \, \mu , \, \boldsymbol P} =
\sum\limits_{\substack{\vecb{k}_{12}, \, \vecb{k}_3, \, \mu_{12} \\ \vecb{k}_{12}+\vecb{k}_3 = \vecb{P}}}
\mathcal{C}(\vecb{P}, \Lambda, \mu; \vecb{k}_{12}, \Lambda_{12}, \mu_{12}; \vecb{k}_3, \Lambda_3) \;
(\boldsymbol \pi \boldsymbol \pi)^{[\vecb{k}_1,\vecb{k}_2]\dagger}_{I_{2}, \Lambda_{12}, \, \mu_{12}}(\vecb{k}_{12}) \cdot \boldsymbol C_{\sf iso}(2, I_2, 1) \cdot \boldsymbol \pi^\dagger_{\Lambda_3}(\vecb{k}_3) \, ,
\label{equ:3piops}
\end{equation}
where the sum is over all momenta related to $\vecb{k}_{12}$ and $\vecb{k}_3$ by allowed lattice rotations and, again, the isospin projection $1 \otimes I_2 \to I_3=2$ is represented with implicit indices. From Bose symmetry, a three-pion system must be symmetric under the interchange of any pair of pions. Since pions have no intrinsic spin, the flavour and spatial structure must therefore be symmetric under interchange of any pair of pions. As discussed in Ref.~\cite{Hansen:2020otl}, Eq.~\eqref{equ:3piops} does not make this symmetry manifest and two different sets of ($|\vecb{k}_1|$, $|\vecb{k}_2|$, $|\vecb{k}_3|$, $|\vecb{k}_{12}|$, $I_{2}$, $\Lambda_{12}$) may lead to equivalent operators or several different sets may give linearly-dependent operators.\footnote{Rather than magnitudes of momentum, strictly this should be types of momenta (i.e.\ momenta related by rotations in the octahedral group or little group), but there is no distinction for the momenta we are considering here.}

To ensure we have an appropriate set of independent operators,
we generalize the approach used in Ref.~\cite{Hansen:2020otl} to $I_3=2$. The essential idea is to represent the combined Clebsch-Gordan coefficients needed to project both to a given total isospin and a given finite-volume irrep as a single vector, denoted by $\widetilde {\boldsymbol {\mathcal C}}$, and defined to satisfy
\begin{equation}
(\boldsymbol \pi \boldsymbol \pi \boldsymbol\pi) = \widetilde {\boldsymbol {\mathcal C}} \cdot \boldsymbol {\mathcal V}_{[\pi \pi \pi]} \,,
\end{equation}
where $(\boldsymbol \pi \boldsymbol \pi \boldsymbol \pi)$ represents any three-pion operator as constructed above and $\boldsymbol{\mathcal V}_{\pi \pi \pi}$ is a vector populated with operators of the form $\pi^\dagger(\vecb{p}_1, I_{z_1}) \pi^\dagger(\vecb{p}_2, I_{z_2}) \pi^\dagger(\vecb{p}_3, I_{z_3})$, i.e.~three-pion operators with definite individual momenta and isospins.

The next step is to perform a symmetrization operation on this vector to account for the exchange symmetry of pions
\begin{equation}
\widetilde {\boldsymbol {\mathcal C}}^{\mathcal S} = \widetilde {\boldsymbol {\mathcal C}} \cdot {\boldsymbol {\mathcal S}} \,, \quad \text{where} \quad {\boldsymbol {\mathcal S}} = \sum_{\text{perm}(ijk)\ {\rm of}\ 123} \boldsymbol R(ijk) \,,
\end{equation}
with the sum running over the six permutations of a three element set. Here $\boldsymbol R(ijk)$ is defined by its action on $\boldsymbol {\mathcal V}_{[\pi \pi \pi]}$, in particular by mapping a given entry $\pi^\dagger(\vecb{p}_1, I_{z_1}) \pi^\dagger(\vecb{p}_2, I_{z_2}) \pi^\dagger(\vecb{p}_3, I_{z_3})$ into $\pi^\dagger(\vecb{p}_i, I_{z_i}) \pi^\dagger(\vecb{p}_j, I_{z_j}) \pi^\dagger(\vecb{p}_k, I_{z_k})$.
We have performed this construction of $ \widetilde {\boldsymbol {\mathcal C}}^{\mathcal S}$ for every possible set ($|\vecb{k}_1|$, $|\vecb{k}_2|$, $|\vecb{k}_3|$, $|\vecb{k}_{12}|$, $I_{2}$, $\Lambda_{12}$) with $|\vecb{k}_1|^2 + |\vecb{k}_2|^2 + |\vecb{k}_3|^2$ less than some cutoff and have then identified a linearly independent set of the resulting vectors corresponding to a non-redundant operator set.

As discussed in Ref.~\cite{Woss:2019hse}, if we were to only use these $\pi\pi\pi$ operators, we would be trying to describe the eigenstates of the $\pi\pi$ subsystem using only $\pi\pi$ operators. However, to reliably determine spectra in the isovector $\pi\pi$ channel (where the $\rho$ resonance appears), one should also include fermion-bilinear operators~\cite{Wilson:2015dqa}. Therefore, we follow Ref.~\cite{Cheung:2017tnt} and use $\rho^{\mathfrak{n} \, \dagger}_{\Lambda_i, \, \mu_i}(\vecb{k}_i)$ operators which are the optimal combinations of fermion-bilinear and $\pi\pi$ operators\footnote{The operator bases are described in Ref.~\cite{Dudek:2012xn}}
to interpolate the $\mathfrak{n}$'th state in irrep $\Lambda_i$ with momentum $\vecb{k}_i$. Following Eq.~(5) of Ref.~\cite{Woss:2019hse}, the $\rho\pi$ operators are then,
\begin{equation}
(\boldsymbol \rho \boldsymbol \pi)^{[I_3=2, \Lambda_{12};\vecb{k}_{12},\vecb{k}_3]\dagger}_{\Lambda, \, \mu}(\vecb{P})=
\sum\limits_{\substack{\vecb{k}_{12}, \, \vecb{k}_3, \, \mu_{12} \\ \vecb{k}_{12}+\vecb{k}_3 = \vecb{P}}}
\mathcal{C}(\vecb{P}, \Lambda, \mu; \vecb{k}_{12}, \Lambda_{12}, \mu_{12}; \vecb{k}_3, \Lambda_3) \;
\boldsymbol \rho^{\mathfrak{n} \, \dagger}_{\Lambda_{12}, \, \mu_{12}}(\vecb{k}_{12}) \cdot \boldsymbol C_{\sf iso}(I_3=2, 1, 1) \cdot \boldsymbol \pi^\dagger_{\Lambda_3}(\vecb{k}_3) \, ,
\label{equ:rhopiops}
\end{equation}
where again we have suppressed explicit isospin indices.

\begin{table}[h!]
\centering
\renewcommand{\arraystretch}{1.3}
\begin{tabular}{lccc}
\hline
Ops. & $~~~a_t E^{\sf non{\text{-}}int}(L/a_s = 16)~~~$ & $~~~a_t E^{\sf non{\text{-}}int}(L/a_s = 20)~~~$ & $~~~a_t E^{\sf non{\text{-}}int}(L/a_s = 24)~~~$\\
\hline
$ [(\pi_{[001]})_{A_2} (\pi_{[00-1]})_{A_2}]_{T_1^-} \, (\pi_{[000]})_{A_1^-}$ & -- & -- & 0.275\\
$ [(\pi_{[011]})_{A_2} (\pi_{[0-1-1]})_{A_2}]_{T_1^-}(\pi_{[000]})_{A_1^-}$ & -- & -- & 0.325 \\
$[(\pi_{[-101]})_{A_2} (\pi_{[100]})_{A_2}]_{A_1} (\pi_{[00-1]})_{A_2} $ & -- & -- & 0.333 \\
$[(\pi_{[-101]})_{A_2} (\pi_{[100]})_{A_2}]_{E_2} (\pi_{[00-1]})_{A_2} $ & -- & -- & 0.333 \\
$[(\pi_{[111]})_{A_2} (\pi_{[000]})_{A_1^-}]_{A_1} (\pi_{[-1-1-1]})_{A_2} $ & -- & -- & 0.366 \\
$[(\pi_{[002]})_{A_2} (\pi_{[00-1]})_{A_2}]_{A_1} (\pi_{[00-1]})_{A_2} $ & -- & -- & 0.372 \\
\hline
$(\rho^{\mathfrak{n}=0}_{[000]})_{T_1^-} (\pi_{[000]})_{A_1^-}$ & 0.225 & 0.223 & 0.220 \\
$
(\rho^{\mathfrak{n}=0}_{[001]})_{A_1} (\pi_{[00-1]})_{A_2}$ & 0.315 & 0.286 & 0.267 \\
$
(\rho^{\mathfrak{n}=0}_{[001]})_{E2} (\pi_{[00-1]})_{A_2}$ & 0.328 & 0.294 & 0.274 \\
$
(\rho^{\mathfrak{n}=1}_{[001]})_{A_1} (\pi_{[00-1]})_{A_2}$ & 0.346 & 0.307 & 0.282 \\
$(\rho^{\mathfrak{n}=0}_{[011]})_{A_1} (\pi_{[0-1-1]})_{A_2}$ & 0.393 & 0.343 & 0.311\\
$(\rho^{\mathfrak{n}=0}_{[011]})_{B_1} (\pi_{[0-1-1]})_{A_2}$ & 0.398 & 0.346 & 0.313\\
$(\rho^{\mathfrak{n}=0}_{[011]})_{B_2} (\pi_{[0-1-1]})_{A_2}$ & 0.400 & 0.348 & 0.314
\\

\hline
\end{tabular}
\renewcommand{\arraystretch}{1.}
\caption{
Summary of operators used for the different lattice volumes.
The corresponding non-interacting energies are shown for operators that were used in the calculation, while -- indicates that the operator was not used on that lattice volume.
As emphasized in the text, the $\rho^{\mathfrak{n}}$ operators are optimized combinations of fermion-bilinear and $\pi\pi$ operators and are sufficient, by themselves, to reliably determine the spectrum for the smaller volumes.
}
\label{tab:operators}
\end{table}

\begin{figure*}[t]
\begin{center}
\includegraphics[width=.85\textwidth]{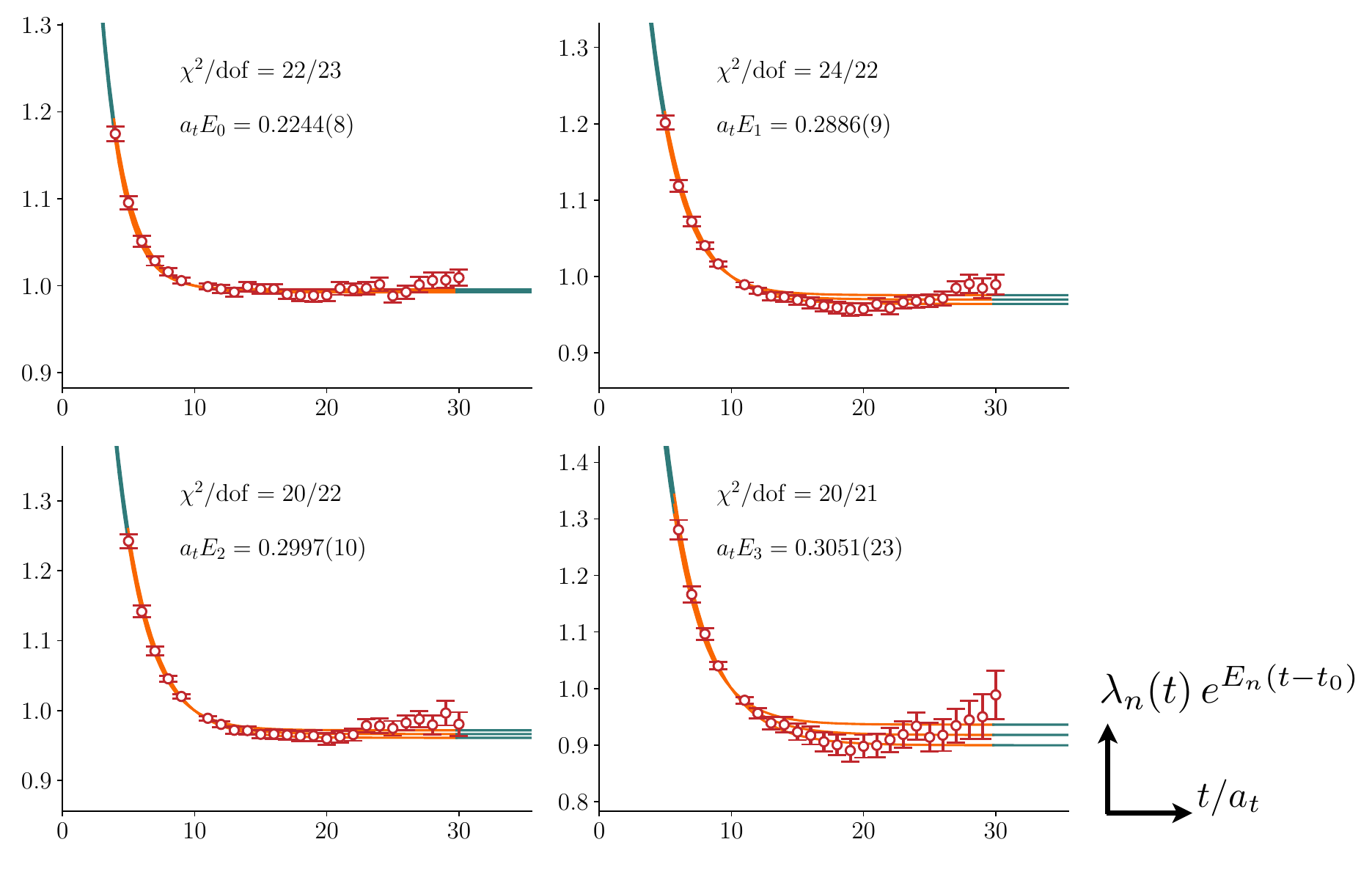}
\caption{
\label{fig:prin_corr}
The four lowest eigenvalues on the $L/a_s = 20$ volume, scaled by $e^{E_n(t-t_0)}$, where
$t_0 = 10$ was used in this case. On each plot, the curves show the result of a two-exponential fit to the highlighted time region as described in the text, and the fitted energy $E_n$ and goodness of fit are also indicated.
}
\end{center}
\end{figure*}

After calculating a Hermitian matrix of correlation functions $\textbf{C}(t)$, with elements $C_{ij}(t) = \langle 0 | \mathcal{O}_i^{\vphantom{\dagger}}(t) \mathcal{O}_j^\dagger(0) | 0 \rangle $,
on each lattice volume using the operators $\{ \mathcal{O}_i \}$ listed in Table \ref{tab:operators}, we solve the generalized eigenvalue problem~\cite{Michael:1985ne,Luscher:1990ck,Blossier:2009kd}\begin{align}
\textbf{C}(t) \cdot \vec{v}^{\,(n)}(t,t_0) = \lambda_n(t,t_0) \,
\textbf{C}(t_0) \cdot \vec{v}^{\,(n)}(t,t_0),
\end{align} where $t_0$ is a time where the eigenvalues, $\lambda_n(t,t_0)$, are fixed to be $1$ for all eigenvectors $\vec{v}^{\,(n)}$. For large times the eigenvalues behave as $\lambda_n(t,t_0) \sim e^{- E_n(t-t_0)}$, where $E_n$ is the energy of the nth state.
Our procedure to solve the generalized eigenvalue problem, obtain the eigenvalues and eigenvectors, is outlined in Ref.~\cite{Dudek:2010wm}.
As discussed there, we fit the eigenvalues to the form,
\begin{equation}
\lambda_n(t,t_0) = (1 - A_n) e^{-E_n (t - t_0)} + A_n e^{-E_n' (t - t_0)} \, ,
\end{equation}
where the fit parameters are $E_n$, $E_n'$, and $A_n$, and the second exponential is used to account for some residual excited state contamination.
As an example of the result of this procedure, Fig.~\ref{fig:prin_corr} shows the four lowest eigenvalues obtained on the $L/a_s = 20$ volume, scaled by $e^{ E_n(t-t_0)}$.

\section{Details of the spectrum analysis}
\label{app:spec_anal}

We begin by specifying the quantization condition. In this work, we adopt the asymmetric formalism, in which the intermediate three-pion K matrix $\textbf K_3$ does not obey the same exchange symmetry as the three-pion scattering amplitude~\cite{Jackura:2022gib}. The choice between the symmetric and asymmetric formalisms reflects a scheme dependence. In particular, the relation between the finite-volume energies and $\textbf K_3$ is scheme dependent, as is $\textbf K_3$ itself. But crucially, the relation between $\textbf K_3$ and the scattering amplitude again has this same scheme dependence, in such a way that the scattering amplitude is the universal quantity that can be directly related to experimental observables.

In the asymmetric formalism, the finite-volume energies satisfy
\begin{align}
\det_{I_2,\boldsymbol{k},S,m}\left[ \textbf F_3^{-1}(E, L \, \vert \, \boldsymbol \eta_2) + \textbf K_3(E \, \vert \, \boldsymbol \eta_3) \right] = 0 \,,
\end{align}
where $E$ is simultaneously the center-of-momentum frame (CMF) energy and the energy in the finite-volume frame, since we only consider the case of vanishing total spatial momentum.
As indicated, the determinant is taken over the direct product space spanned by the isospin of the two-pion subchannel $I_2$, the spectator momentum $\boldsymbol{k}$, the total angular momentum $S$ of the two-pion channel, and the corresponding magnetic quantum number $m$.
The two main objects that appear are $\textbf{K}_3$, the asymmetric three-pion K matrix for total isospin two (discussed further in Sec.~\ref{app:K3_constrain}), and $\textbf{F}_3$, which depends on the finite-volume periodicity $L$ as well as the total energy $E$. In addition, $\textbf{F}_3$ depends on the two-to-two scattering amplitudes for all two-pion subchannels that can contribute for this total isospin. We use the notation $\boldsymbol{\eta}_2$ to denote the parameters that describe these subprocesses.

In the asymmetric formulation that we use here, the $\textbf F_3$ matrix is given by
\begin{equation}
\textbf F_3(E, L \, \vert \, \boldsymbol \eta_2) = \frac{1}{L^3} \Big [ \Big ( \textbf F_2(E, L) + \textbf G(E, L) \Big )^{-1}+ \textbf K_2(E \, \vert \, \boldsymbol \eta_2) \Big ]^{-1} \,,
\label{eq:F_3}
\end{equation}
where all boldfaced building blocks on the right-hand side are matrices in the $I_2, \boldsymbol k,S,m$ space. $\textbf{F}_2$ encodes finite-volume effects from two-pion subprocess scattering without exchanges, $\textbf{G}$ accounts for such exchanges, and $\textbf{K}_2$ depends on the two-pion scattering parameters.

The explicit construction of these matrices follows Refs.~\cite{Romero-Lopez:2019qrt,Alotaibi:2025pxz} with two important subtleties. First, we use the standard definition of spherical harmonic, $Y_{S m}(\hat{\vecb{k}}) =\langle \hat{\vecb{k}} |S m\rangle$. This affects which spherical harmonics are conjugated inside of the finite-volume functions. Second, here we make barrier factors explicit, whereas in Refs.~\cite{Romero-Lopez:2019qrt,Alotaibi:2025pxz} these are removed by rescaling definitions in a way leaves the quantization condition unchanged.

We now proceed to review the key expressions needed for this work, beginning with $\textbf{K}_2$, whose matrix elements are given by
\begin{equation}
\left[\textbf K_2\right]_{ {I_2'}{\boldsymbol k'}{S'}{m'} , {I_2}{\boldsymbol k}{S}{m} }
= \delta_{I_2'I_2}\delta_{{S'}S}\delta_{{m'}m}\delta_{{\boldsymbol k'}\boldsymbol k} \frac{1}{2 \omega_k} \mathcal K_2^{I_2 S}(\sigma_k) \,,
\end{equation}
where $\Kc^{I_2 S}_2(\sigma_k)$ is a scheme-dependent two-body K matrix and $\sigma_k$ is the squared CMF energy of the two-pion subsystem, which is the same as $M_{\pi\pi}^2$. The two relevant components are
\begin{align}
\Kc^{I_2=2,S=0}_2(\sigma_k)^{-1} & = \frac{q^{\star}_k \cot \delta_{2}(\sigma_k) + \vert q^\star_k \vert [1 - J(z_2(\sigma_k))]}{16 \pi \sqrt{\sigma_k}} \,, \\
\Kc^{I_2=1,S=1}_2(\sigma_k)^{-1} & = \frac{1}{16 \pi \sqrt{\sigma_k}} \left [ q^{\star}_k \cot \delta_{1}(\sigma_k) + \vert q^\star_k \vert [1 - J(z_1(\sigma_k))] - \Ic_{\rm PV}\frac{ \sqrt{\sigma_k} m_\pi^2}{2 q_k^{\star 2} } \right ] \,.
\label{eq:K2_BW}
\end{align}
Here $\delta_{I_2}(\sigma_k)$ is the scattering phase shift for the $I_{2}$ channel, and $q^{\star}_k$ is the relative momentum of the two-particle subsystem defined by carrying total momentum $P-k$, where $P = (E, \boldsymbol 0)$ is the total four-momentum of the system and $k = (\omega_k, \boldsymbol k)$ is the four momentum of an on-shell pion carrying momentum $\boldsymbol k$.
This information is formally enough to specify all kinematics and implies
\begin{equation}
\sigma_k = (E - \omega_k)^2 - \boldsymbol k^2 \,, \qquad \omega_k = \sqrt{m_\pi^2 + \boldsymbol k^2} \,, \qquad {q_k^\star}= \sqrt{\sigma_k/4 - m_\pi^2}\,.\label{eq:qk}
\end{equation}

These K matrices are scheme dependent due to the presence of the cut-off function $J(z_{I_2}(\sigma_k))$ and, in the $I_2=1$ case, also due to the shift labeled with $\mathcal{I}_{\rm PV}$. This shift, which we discuss further below and in Sec.~\ref{app:amp_recon}, is a simple modification to the phase space. The cutoff function was first introduced in Ref.~\cite{Hansen:2014eka} and later generalized in Ref.~\cite{Briceno:2017tce}, and is defined as
\begin{align}
\label{eq:J_cutoff1}
J(z) & =
\begin{dcases}
0, & 0 \geq z \,,\\
\exp\!\left[-\frac{1}{z} \exp\left(-\frac{1}{1-z}\right)\right], & 0 < z < 1 \,,\\
1, & 1 \leq z \,,
\end{dcases}
\\
\label{eq:J_cutoff2}
z_{I_2}(\sigma_k)&=
\frac{{\sigma_k}/(4 m_\pi^2) - (1+\alpha_{I_2})/4}{(3-\alpha_{I_2})/4}.
\end{align}
The advantage of the form introduced in Ref.~\cite{Briceno:2017tce}, which we adopt here, is the introduction of a parameter, $\alpha_{I_2}$, that controls the location of the zero of the cutoff function. In particular, the cut-off function vanishes for $\sigma_k \leq (1+\alpha_{I_2})m_\pi^2$. This flexibility is crucial for eliminating subthreshold singularities that would otherwise invalidate the formalism, as we discuss further below.
For the $I_2=1$ channel we have introduced an additional parameter, $\mathcal{I}_{\rm PV}$, which shifts the inverse K matrix as shown. This was proposed in Ref.~\cite{Romero-Lopez:2019qrt} to treat the case of a physical pole in the K matrix, in the scattering region, which would otherwise invalidate the formalism.

\begin{figure*}[t]
\begin{center}
\includegraphics[width=.9\textwidth]{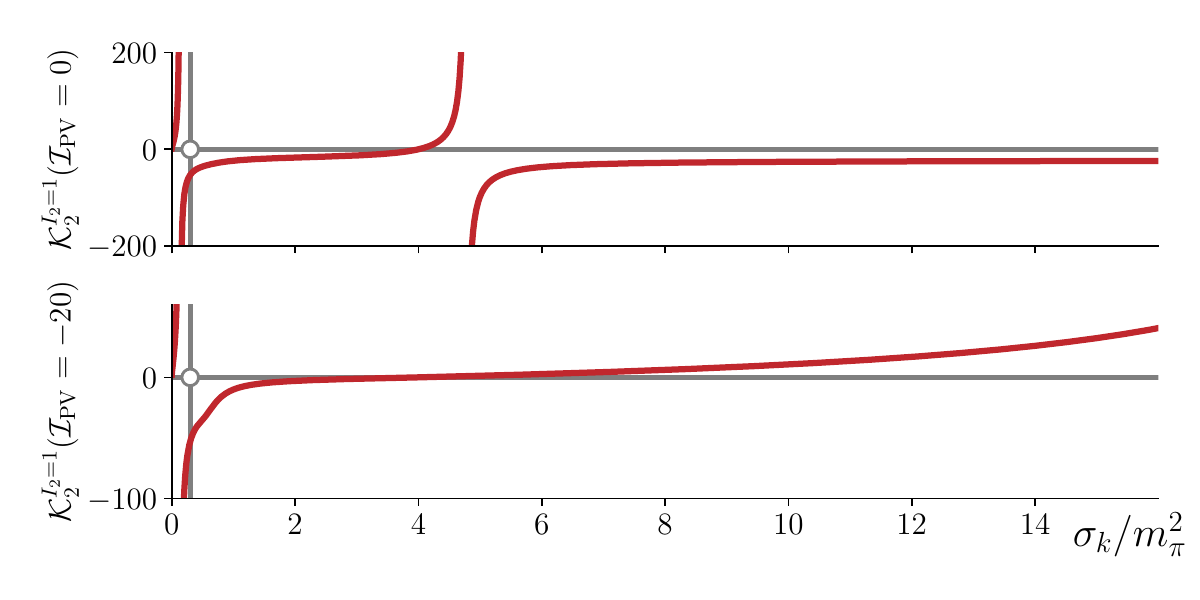}
\caption{Shown are the values of $\Kc^{I_2=1}_2$ using $m_0/m_\pi=2.184$ and $g_0=5.80$ for the Breit-Wigner parameters. The top and bottom panels show this function before and after the shift ($\Ic_{\rm PV}$) appearing in Eq.~\ref{eq:K2_BW}. The vertical gray line, defined by $\sigma_k=(1+\alpha_{1})=0.3 m_\pi^2$, highlights the point where the cut off function is fixed to $0$.\label{fig:Pwave_K2}
}
\end{center}
\end{figure*}

In this work, we use the leading order effective range expansion to parametrize the $I_2=2$ channel,
\begin{equation}
q^{\star}_k \cot \delta_{2}(\sigma_k) = - \frac{1}{a_2} \,,
\end{equation}
where $a_2$ is the isotensor scattering length. This has been previously determined in Refs.~\cite{Dudek:2012gj, Hansen:2020otl}. We fix this parameter to its central value, which we take from Ref.~\cite{Hansen:2020otl} to be $a_2 m_\pi = 0.296(8)$.
For the $I_2=1$ channel, we use the Breit-Wigner parametrization,
\begin{align}
q^{\star 3}_k \cot \delta_{1}(\sigma_k) & = \frac{6\pi}{g_0^2} \sqrt{\sigma_k} (m_0^2 - \sigma_k) \,,
\end{align}
where $m_0$ is Breit-Wigner mass and $g_0$ is the coupling of the resonance to the two-pion channel. Again, these have been previously constrained in Ref.~\cite{Dudek:2012xn} to be $m_0/m_\pi=2.184(49)$ and $g_0=5.80(10)$ with a correlation of $-0.26$ for these same ensembles. Again, we fix the parameters to their central values from this previous work.

Using these Breit-Wigner parameters with $\alpha_1 = -1$ and $\Ic_{\rm PV} = 0$, $\Kc^{I_2=1,S=1}_2(\sigma_k)$ has two poles that invalidate the formalism. One is a subthreshold pole located at $\sigma_k \approx 0.134 \, m_\pi^2$. This is removed by setting $\alpha_1 = -0.7$, rather than $-1$.
The other pole is located at $\sigma_k=m_0^2$ and is removed by taking a nonzero value of $\Ic_{\rm PV} = -20$.
This modification of the two-particle K matrices must be accompanied by a modification of the integral equations relating the phase space appearing in the scattering amplitudes. We discuss this in Sec.~\ref{app:int_eq}.
In Fig.~\ref{fig:Pwave_K2}, we show the value of $\Kc^{I_2=1,S=1}_2(\sigma_k)$ in the kinematic range considered for $\Ic_{\rm PV} = 0$ and $-20$. As one sees clearly, before the shift, this function has a pole at $\sigma_k=m_0^2\approx 4.7m_\pi^2$, which is removed by the shifting. This shift only slightly modifies the deeper pole. In both panels, we highlight $\sigma_k=(1+\alpha_1)m_\pi^2=0.3 m_\pi^2$, where the smooth cut off previously discussed is fixed to $0$. Finally, we note that, although $\alpha_2 = -1$ would not lead to any subthreshold singularities in the $I_2=2$ channel for the kinematics considered here, we also set $\alpha_2 = -0.7$ for consistency.

Now we turn our attention to the $\textbf{F}_2$ matrix, which has matrix elements of the form,
\begin{align}
\left[\textbf{F}_2\right]_{ {I_2'}{\boldsymbol k'}{S'}{m'} ; {I_2}{\boldsymbol k}{S}{m} }
&=
\delta_{I_2' I_2}\delta_{\boldsymbol k' \boldsymbol k}
\frac{J(z_{I_2}(\sigma_{k}))}{2\omega_k}
\left[
\left[F(\boldsymbol k)\right]_{{S'} {m'};S m}
+ \delta_{I_2,1}\delta_{S',1}\delta_{S,1}\delta_{m'm}
\Ic_{\rm PV}\frac{m_\pi^2}{32 \pi q_k^{\star 2}}
\right] \,,
\end{align}
where $F$ is the standard two-body $F$ function that appears in the literature, e.g.~Ref.~\cite{Kim:2005gf}, which the important distinction of which spherical harmonic is conjugated inside the summand/integrand
\begin{equation}
\label{eq:Fscdef}
\left[F(\boldsymbol k)\right]_{{S'} {m'};S m} = \frac{1}{2}
\bigg[\frac{1}{L^3}\sum_{\vecb{a}} - \int\frac{d^3\vecb{a}}{(2\pi)^3} \bigg]
\frac{ 4 \pi Y_{S'm'}^*(\hat {\vecb a}^\star_k)Y_{S m}(\hat {\vecb a}^\star_k) }{2 \omega_{a} [(P-k-a)^2-m_\pi^2 + i \epsilon]} \left (\frac{\vert \boldsymbol a^{\star}_k \vert}{q^\star_k} \right )^{S+S'} \,,
\end{equation}
where $a = (\omega_a, \vecb{a})$. Here we have also introduced $a_k^\star$ as the four vector given by boosting $a$ with boost velocity $\boldsymbol \beta_k = \boldsymbol k/(E - \omega_k)$, i.e.~boosting to the CMF of the non-$k$ pair (sometimes called the dipion). Then $\boldsymbol a^\star_k$ is the spatial part of $a_k^\star$ and $\hat {\vecb a}^\star_k = {\vecb a}^\star_k/\vert {\vecb a}^\star_k \vert$ is its direction.

Finally, the matrix elements of $\textbf{G}$ can be written as,
\begin{align}
\left[\textbf{G}\right]_{ {I_2'}{\boldsymbol{k}'}{S'}{m'} ; {I_2}{\boldsymbol{k}}{S}{m} }
=\Rc_{{I_2'}{I_2}} J(z_{I_2'}(\sigma_{k'}))
\overline G_{ {\boldsymbol{k}'}{S'}{m'} ; {\boldsymbol{k}}{S}{m} }
J(z_{I_2}(\sigma_{k})) \,,
\end{align}
where $\overline G$ is the exchange propagator with spherical harmonics, barrier factors, and a normalization factor
\begin{align}
\overline G_{ {\boldsymbol{k}'}{S'}{m'} ; \,{\boldsymbol{k}}{S}{m} }
=
\frac{1}{2\omega_{k'} 2 \omega_{k} L^3}
\frac{
\vert \vecb{k}'^\star_{k}\vert^{S'}
\vert \vecb{k}^\star_{k'}\vert^S}
{\left({q}^\star_{k}\right)^{S'}
\, \left({q}^\star_{k'}\right)^{S}}
\frac{4\pi Y^*_{S 'm'}(\hat{\vecb{k}}'^\star_k) Y_{Sm}(\hat{\vecb{k}}^\star_{k'})}{ (P-k'-k)^2 - m_\pi^2 }
\,,
\end{align}
and $\Rc_{{I_2'} {I_2}}$ are isospin recoupling coefficients~\cite{Romero-Lopez:2019qrt,Alotaibi:2025pxz},
\begin{align}
\Rc_{2 2} =-\frac{1}{2}
\,, \qquad
\Rc_{2 1} =\Rc_{1 2} = -\frac{\sqrt{3}}{2}
\,, \qquad
\Rc_{1 1} =\frac{1}{2}.
\end{align}

\section{Subduction to the $T_1^+$ Irrep}
\label{app:T1irrepQC}

The method for projecting $\textbf{F}_3$ onto a particular irreducible representation (irrep) was first presented in Ref.~\cite{Blanton:2019igq}. A variant yielding smaller matrices was later introduced in Ref.~\cite{Alotaibi:2025pxz}, which is the approach adopted in this work. In this appendix, we briefly discuss both methods. They differ only in the specific coefficients used in the projection, as described at the end of this section.

The projection of $\textbf{F}_3$ is achieved in practice by projecting each of its three fundamental building blocks: $\textbf{K}_2$, $\textbf{F}_2$, and $\textbf{G}$. As described in the previous section, each of these matrices is defined in $I_2 \boldsymbol{k} S m$-space, but for the purposes of projecting to a definite finite-volume irrep, the isospin index is irrelevant. We therefore restrict attention to the remaining set and denote the generic object to be projected as $M_{\boldsymbol{k}' S' m',\, \boldsymbol{k} S m}$. To restrict this to a definite cubic irrep, in particular to the $T_1^+$, we must consider its transformations under the elements of the 24-element octahedral group ${\rm O}_h$, together with spatial inversions which we incorporate explicitly.

Begin by noting that one can identify $24$ triplets of Euler angles, $\alpha, \beta, \gamma$, corresponding to the $24$ proper rotations of the octahedral group. Then $\Rc(\alpha, \beta, \gamma)$ denotes a generic real three-dimensional spatial rotation matrix defined by such Euler angles, and thereby provides a representation of ${\rm O}_h$. For any such $\Rc$, together with a possible spatial inversion $\pi \in \{0, 1\}$, we define the transformation of a momentum vector $\vecb{k}$ by
\begin{align}
\vecb{k} \ \to \ \vecb{k}' = (-1)^\pi \Rc(\alpha, \beta, \gamma) \cdot \vecb{k} \,.
\end{align}

We next consider the action of such a transformation on the $S m$ indices of $M$. This is inherited from its action on the spherical harmonic $Y_{S m}(\hat{\vecb{k}})$, which in turn follows from the fact that a rotation of the harmonic is equivalent to a rotation of the coordinate system in the opposite direction. Beginning with proper rotations, we write
\begin{align}
Y_{S m'}(\hat{\vecb{k}})
\to
Y_{S m'}(\Rc^T(\alpha,\beta,\gamma)\cdot \hat{\vecb{k}})
&=
\langle \Rc^T(\alpha,\beta,\gamma) \cdot \hat{\vecb{k}} |S m'\rangle
\\
&=
\langle \hat{\vecb{k}} |\,\Rc(\alpha,\beta,\gamma) \,|S m'\rangle
\\
&=
\sum_{m}
\langle \hat{\vecb{k}} |S m\rangle \langle S m | \,\Rc(\alpha,\beta,\gamma) \,|S m'\rangle
\\
&=
\sum_{m}
D_{mm'}^{(S)}(\alpha,\beta,\gamma)
\,Y_{S m}(\hat{\vecb{k}}),
\end{align}
where $D^{(S)}$ is the Wigner D matrix. Its matrix elements are defined as,
\begin{align}
D_{m' m}^{ ( S ) }(\alpha, \beta, \gamma )
& = \langle S m' | \Rc(\alpha, \beta, \gamma) | S m \rangle = e^{-im' \alpha} \, d_{m'm}^{(S)}(\beta) \, e^{-im \gamma} \, ,
\end{align}
where $ d^{(S)}$ is the reduced Wigner D matrix, defined for example in Ref.~\cite{VMK}. From this, it follows that a proper rotation of the spherical harmonic takes the form
\begin{align}
Y_{S m'}^*(\hat{\vecb{k}})
\to
\sum_{m}
D_{mm'}^{(S)*}(\alpha,\beta,\gamma)
\,Y_{S m}^*(\hat{\vecb{k}}).
\end{align}

We are now in a position to determine the transformation of $M_{\boldsymbol k'S'm',\boldsymbol kS m}$ under a rotation, $\alpha, \beta, \gamma$, and parity transformation, $\pi$. Following Ref.~\cite{Blanton:2019igq}, we write the transformation in terms of two matrices, $S$ and $W$, defined by
\begin{align}
S_{\boldsymbol k'S'm', \boldsymbol kS m }(\alpha, \beta, \gamma, \pi) &=
\begin{cases}
\delta_{S' S}\delta_{m m'},& \text{if } \vecb{k}' =(-1)^\pi\Rc(\alpha,\beta,\gamma) \cdot \vecb{k} \,;\\
0, & \text{otherwise} \,,
\end{cases}
\\[5pt]
W_{\boldsymbol k'S'm',\boldsymbol kS m }(\alpha, \beta, \gamma, \pi) &=
(-1)^{\pi S} \,\delta_{S' S}\,
\delta_{k' k}\,
D_{m m'}^{ ( S ) * }(\alpha, \beta, \gamma ) .
\end{align}
To make the notation more compact, in the following we represent a given $\alpha, \beta, \gamma$ by the matrix $R = \mathcal R(\alpha, \beta, \gamma)$.
With this, we can define the transformation of $M$ as
\begin{align}
M_{\boldsymbol k'S'm', \boldsymbol kSm}
\rightarrow \left[W(R, \pi) \cdot S(R, \pi)\cdot M \cdot S^T(R, \pi) \cdot W^\dag(R, \pi) \right]_{\boldsymbol k'S'm', \boldsymbol kSm} \,.
\end{align}
Given that $S$ and $W$ commute, the order of multiplication here is arbitrary.

We now define the projector. The essential idea is to form an appropriately weighted sum over all cubic rotations and inversions. Denoting the weights by $w_{\Lambda}(R,\pi)$, we define
\begin{align}
P_{\Lambda} = \sum_{\pi =0,1} \sum_{R \in \mathrm{O}_h} \, w_{\Lambda}(R, \pi)\, W(R, \pi) \cdot S(R, \pi) \,.
\label{eq:P_Lambda_def}
\end{align}
An important point is that this is a standard projector, satisfying $P_\Lambda \cdot P_\Lambda = P_\Lambda$ and, in particular, it is a square matrix. Thus, to actually shrink the size of generic matrices of the form $M_{\boldsymbol{k}' S' m',\, \boldsymbol{k} S m}$, we need to identify the non-trivial subspace of the projector. To achieve this, we determine its eigenvalues and eigenvectors
\begin{align}
P_{\Lambda} \cdot \hat {\boldsymbol e}_n
=\lambda_n \hat {\boldsymbol e}_n \,.
\end{align}
As $P_\Lambda$ is generally not a symmetric matrix, one must be careful as to ensure that these are the \emph{right eigenvectors}. Also, because these are not symmetric matrices, and because the non-zero eigenvalues are degenerate, one must ensure that the basis is orthogonal. We do so using the standard Gram-Schmidt process.

Finally, we construct a projector that reduces the size of all building blocks by retaining only the subset of eigenvectors with non-zero eigenvalues. Denoting this subset by $\left\{\hat{\boldsymbol{e}}_1, \cdots, \hat{\boldsymbol{e}}_N\right\}$, we write
\begin{align}
\widetilde{P}_{\Lambda}^\dagger \equiv
\begin{pmatrix}
\hat {\boldsymbol e}_1 & \hat {\boldsymbol e}_2 & \cdots & \hat {\boldsymbol e}_N
\end{pmatrix} \,.
\end{align}
Given this, we finally define the projected $M$ as
\begin{align}
M_{\Lambda}
\equiv
\widetilde{P}_{\Lambda}
\cdot
M\cdot
\widetilde{P}_{\Lambda}^\dag \,.
\end{align}

At this stage, it remains only to define the weights $w_{\Lambda}(R, \pi)$. We describe two options, starting with that described in Ref.~\cite{Blanton:2019igq}.
We first introduce $\mathfrak{D}^{(\Lambda)}(R, \pi)$ as a $n_\Lambda \times n_\Lambda$ matrix for the irrep of a given group element in some fixed orthonormal basis. We then let $\chi_\Lambda(R, \pi) = \mathrm{tr}\, \mathfrak{D}^{(\Lambda)}(R, \pi)$ denote its character.
Then, we can define the standard character projector by setting
\begin{equation}
w_{\Lambda}(R, \pi) = \frac{n_\Lambda}{48} \chi_\Lambda(R^{-1}, \pi) \,,
\end{equation}
where we have used that the group has 48 elements. For unitary $\Lambda$ we can equivalently write,
\begin{equation}
w_{\Lambda}(R, \pi) = \frac{n_\Lambda}{48} \chi_\Lambda(R, \pi)^* \,,
\end{equation}
where the $*$ denotes complex conjugation.
Used in Eq.~\eqref{eq:P_Lambda_def}, this will yield a projector onto all rows of $\Lambda$. So if we let \(m_\Lambda\) be the multiplicity of \(\Lambda\) in \(M\), then $P_\Lambda$ will have $m_\Lambda \times n_\Lambda$ non-zero eigenvalues and $M_\Lambda$ will have this dimension.

The alternative, emphasized in Ref.~\cite{Alotaibi:2025pxz},
aims to remove a redundancy in the projection.
The idea is to define a row-specific projector instead.
We can readily do so by using matrix elements of $\mathfrak{D}^{(\Lambda)}(R, \pi)$ to define
\begin{equation}
w_\Lambda(R, \pi)= \frac{n_\Lambda}{48} \big(\mathfrak{D}^{(\Lambda)}_{\mu\mu}(R^{-1}, \pi)\big) \,,
\end{equation}
where $\mathfrak{D}_{\mu \nu}$ is an element of $\mathfrak{D}$ in the $n_\Lambda \times n_\Lambda$ space and here $\mu$ is fixed to a given value, i.e.~is not summed. Because the $\mu \nu$-matrix elements of a given representation are orthogonal over the summation of the group elements, this is a valid choice of weight associated with a single row. As described in Ref.~\cite{Alotaibi:2025pxz}, using this in the procedure outlined above gives a version of $M_\Lambda$ whose dimension is reduced by a factor of $n_\Lambda$ relative to the alternative.

\section{Constraining $\Kc_3$ from lattice QCD}
\label{app:K3_constrain}
In this section, we discuss details for constraining $\Kc_3$ from the lattice QCD spectra shown in Fig.~\ref{fig:spec}. As is now standard, our strategy is to parametrize $\Kc_3$, subsequently predict the corresponding finite-volume spectrum using Eq.~\eqref{eq:QC}, and define a $\chi^2$ comparison between the predicted spectrum to the lattice QCD spectrum. In practice, many parametrizations are fitted to form systematic error estimates. In the following subsections, we discuss the parametrizations we use for $\Kc_3$, the fitting procedure, and the results for the parameters in each fit.

\subsection{Parametrization}
\label{app:K3_param}

In the quantization condition~\eqref{eq:QC}, the $\Kc_3$ is a matrix in $\vecb{k}Sm$ space (Here we drop the isospin indices for notational convenience), and is a function of the total energy. However, the K matrix of interest and most readily parametrizable is that for definite $J^P$, which is conveniently written in the $\ell S$ basis. We remove angular dependence in the $\vecb{k}Sm$ basis, treating the three-body system as an effective $\2\to\2$ system, where the $\pi\pi$ subsystem is treated as a particle with spin $S$ coupling to the spectator with a relative angular momentum $\ell$. We then expand the K matrix in the $\vecb{k}Sm$ basis by summing over all total angular momenta $J$,
\begin{align}
\label{eq:K3_kSm}
\left[\mathbf{K}_{3}\right]_{\vecb{k}'S'm' , \vecb{k}S m} = \sum_{J = 0}^{\infty} \sum_{\ell' = \lvert J - S'\rvert}^{J+S'} \sum_{\ell = \lvert J-S \rvert}^{J+S}\Kc_{3,\ell'S',\ell S}^{J^P}(k',k) \, \sum_{m_J = -J}^J Z_{\ell'S'}^{Jm_J}(\hat{\vecb{k}}') \,
\left(Z_{\ell S}^{Jm_J}(\hat{\vecb{k}})\right)^*
\end{align}
where we have used conservation of total angular momentum $J$ and projection $m_J$, as well as the Wigner-Eckart theorem to remove the dependence of $m_J$ on the $K$ matrix. The partial-wave K matrix depends implicitly on the total CMF energy, and the magnitudes of the initial and final spectator momenta, $k = \lvert\vecb{k}\rvert$ and $k' = \lvert\vecb{k}'\rvert$, respectively. We introduced the angular $Z$ functions, defined as,
\begin{align}
Z_{\ell S}^{Jm_J}(\hat{\vecb{k}}) = \sqrt{4\pi} \sum_{m_\ell} \, \, \braket{\ell m_\ell,Sm | J m_J} \, Y_{\ell m_\ell}(-\hat{\vecb{k}}) \, .
\end{align}
As discussed in Ref.~\cite{Jackura:2023qtp,Briceno:2024ehy}, the negative sign in the angular function is due to the angular momenta being defined with respect to the pair, which has a momentum $-\vecb{k}$. In this work, we only consider the $J=1$ contribution to the K matrix to be non-zero; thus, the sum over $J$ is restricted to a single term. The remaining K matrix now depends on only three kinematic variables for a given element of the K matrix in the $\ell S$ basis.

Following Ref.~\cite{Briceno:2024ehy}, we parametrize $\Kc_{3;\ell'S',\ell S}^{J^P}(p,k)$ as a separable function of the form,
\begin{align}
\label{eq:K3.separable}
\Kc_{3;\ell'S',\ell S}^{J^P}(p,k) =
\, h_{\ell' S'}(p)
\, \wt{\Kc}_{3;\ell'S',\ell S}^{J^P}(s) \,
h_{\ell S}(k) ,
\end{align}
where $h_{\ell S}(k)$ includes the barrier factors $k^\ell q_k^{\star\,S} /m_\pi^{\ell+S}$ and polynomials in $\sigma_{k}$, or equivalently $ q_k^{\star\,2}$. The reduced K matrix $\wt{\Kc}_3$ is a symmetric matrix, and we have chosen it to be dimensionless. Using spectroscopic notation, $\wt{\Kc}_{3;\ell'S',\ell S}^{1^+} \equiv \wt{\Kc}({}^{2S'+1}\ell'_{1}|{}^{2S+1}\ell_1) $, we write this as a symmetric $3\times 3$ matrix,
\begin{align}
\wt{\Kc}_{3}^{1^+}
=
\left(
\begin{matrix}
\wt{\Kc}({}^3S_1|{}^3S_1) & \wt{\Kc}({}^3S_1|{}^3D_1)& \wt{\Kc}({}^3S_1|{}^1P_1)\\
& \wt{\Kc}({}^3D_1|{}^3D_1) & \wt{\Kc}({}^3D_1|{}^1P_1)
\\
&&\wt{\Kc}({}^1P_1|{}^1P_1) \end{matrix}
\right),
\label{eq:K3_mat}
\end{align}
where we have omitted to show the lower triangular matrix. For physical energies, each component of $\wt{\Kc}_{3}^{J^P} $ must be a real function. In this work, we only consider parametrizations where these are polynomials of $s$ with at most a linear term, i.e., each element has the form
\begin{align}
m_\pi^2 \, \wt{\Kc}({}^{2S'+1}\ell'_{1}|{}^{2S+1}\ell_1) =
\alpha({}^{2S'+1}\ell'_{1}|{}^{2S+1}\ell_1) + \left(\frac{s-(3m_\pi)^2}{m_\pi^2}\right) \, \beta({}^{2S'+1}\ell'_{1}|{}^{2S+1}\ell_1) \,,
\label{eq:K3_1}
\end{align}
where $\alpha({}^{2S'+1}\ell'_{1}|{}^{2S+1}\ell_1)$ and $\beta({}^{2S'+1}\ell'_{1}|{}^{2S+1}\ell_1)$ are dimensionless real constants to be determined. For the $h_{\ell S}(k)$ functions, we consider this to have at most a linear term in $\sigma_k$, i.e.,
\begin{align}
h_{\ell S}(k) = \frac{k^\ell q_k^{\star\,S}}{m_\pi^{\ell+S}}
\left(1+ \gamma({}^{2S+1}\ell_1) \, \frac{q_k^{\star\,2}}{m_\pi^2}\right) \, ,
\label{eq:barriers}
\end{align}
where $\gamma({}^{2S+1}\ell_1)$ is a real constant to be determined.
In the following section, we discuss the fits performed on the spectrum using the parametrizations presented here. An alternative, not considered in this work, might be to use a Bayesian approach along the lines of Ref.~\cite{Salg:2025now}.

\subsection{Fitting procedure}
\label{app:fits}

Here we discuss the fitting of the parameters presented in Supp.~\ref{app:K3_param} to the lattice QCD spectrum shown in Fig.~\ref{fig:spec}. We define a $\chi^2$ between the parametrized finite-volume spectrum and the lattice QCD energies $E_L$, taking into account statistical and two systematic errors in the spectra. The first systematic error is from estimating the variation of the spectrum from different reasonable choices made in fitting the two-point correlation functions (including the $t_0$ and fit range). The second comes from an estimation of exponentially suppressed finite-volume errors that are otherwise ignored. We estimate the latter error by comparing the $m_\pi$ obtained at a fixed finite volume to its infinite-volume extrapolation. This was studied in Ref.~\cite{Dudek:2012gj}, which found the infinite-volume extrapolated mass to be $a_t m_\pi(\infty) = 0.06906(13)$. From this, we introduce an additional error which is equal to $\sigma_L = E_L |m_\pi(L)-m_\pi(\infty)| / m_\pi(\infty)$. In practice, we form ratios of the lattice energies to the finite-volume pion mass, $E_L/m_\pi(L)$, and fit these ratios using the quantization condition with all dimensionful quantities expressed in units of $m_\pi$, i.e. with $m_\pi$ set to one. Adding both systematic errors in quadrature, we then define the covariance matrix used in the fits as $\rm{Cov}=\rm{Cov}_{
\rm stat}+\rm{Cov}_{
\rm syst}$, where $\rm{Cov}_{
\rm stat}$ includes statistical uncertainties and correlations, while $\rm{Cov}_{
\rm syst}$ does not include any correlation across energy levels.

\begin{figure*}[t]
\begin{center}
\includegraphics[width=.8\textwidth]{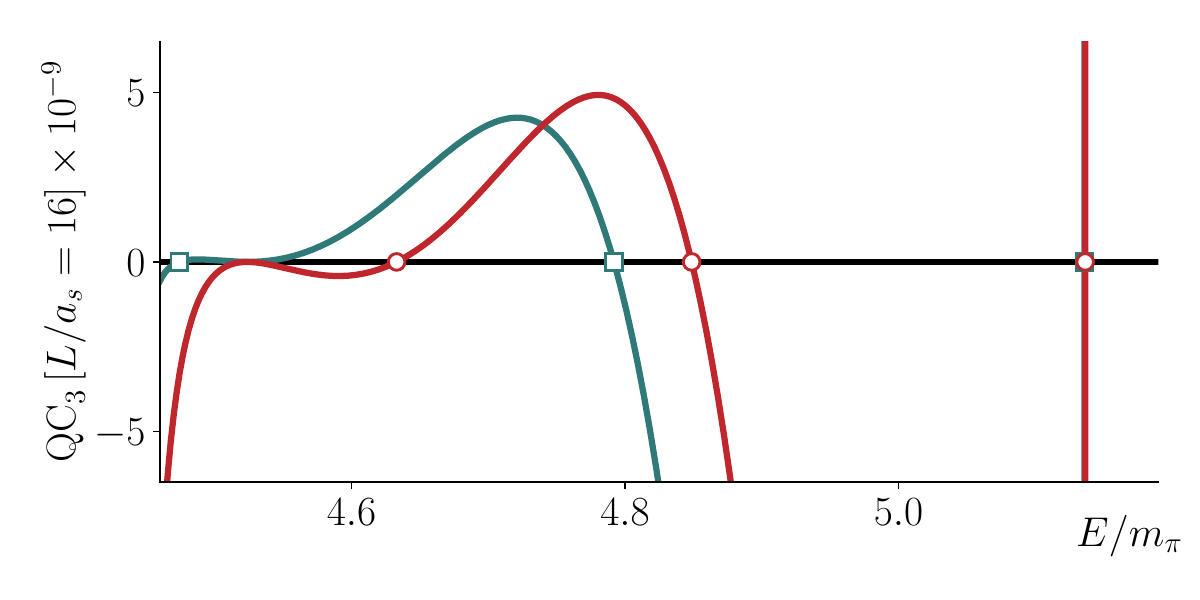}
\caption{
\label{fig:QC3T1_v2}
Shown in red is the value of the ${\rm QC}_3$ function, defined in Eq.~\eqref{eq:QC3_T1}.
Just as in Fig.~\ref{fig:Kmat_amp}, we set the Breit-Wigner parameters to $m_0/m_\pi=2.184$ and $g_0=5.80$ for the Breit-Wigner. The isotensor scattering length is fixed to $a_2 = 0.296/m_\pi$. As discussed in the text, $\Kc_3$ is fixed by a single non-zero parameters, $\alpha({}^3S_1|{}^3S_1) = -9$. The three highlighted zeros appear at $E_L/m_\pi = 4.63, 4.85,5.14$. There is a lower state, not shown, at $E_L/m_\pi = 3.32$. For comparison, we also show the value of this function in the limit that $\Kc_3=0$ in cyan, with its corresponding zeros shown as squares.
}
\end{center}
\end{figure*}
In performing the analysis, we find that the fits can be substantially sped up by pre-computing most of the quantities appearing in the quantization condition. By fixing the two-body parameters to their mean value, one can easily pre-compute $\mathbf{F}_3$, cf. Eq.~\eqref{eq:F_3}, projected to the $T_1^+$ irrep for a large set of nearly continuous energies for the three volumes in our analysis. The three-body K matrix, in general, cannot be pre-computed since this is the quantity we wish to fit. However, for the polynomial parametrizations discussed in Sec.~\ref{app:K3_param}, one can write the $T_1^+$ projected K matrix as a sum of products of kinematic functions that can be pre-computed and coefficients to be fitted.

As an explicit example, consider of subset of the parametrizations defined by Eqs.~\eqref{eq:K3_1} and \eqref{eq:barriers}. In particular, consider the case where all components of the K matrix are zero, except for the ${}^3S_1$ components, which can be written as,
\begin{align}
\wt{\Kc}({}^3S_1|{}^3S_1) =
\alpha({}^3S_1|{}^3S_1) + \left(\frac{s-(3m_\pi)^2}{m_\pi^2}\right) \, \beta({}^3S_1|{}^3S_1),
\label{eq:K3_3S1}
\end{align}
and the barrier factors are the minimal required,
\begin{align}
h_{01}(k) = \frac{q_k^{\star}}{m_\pi}.
\end{align}
Following the steps outlined in this section and Sec.~\ref{app:T1irrepQC}, the $T_1^+$ projected K matrix can be written as,
\begin{align}
\mathbf{K}_{3;T_1^+,n} =
\left[\alpha({}^3S_1|{}^3S_1) +\frac{s-(3m_\pi)^2}{m_\pi^2} \beta({}^3S_1|{}^3S_1) \right]
\wt{\mathbf{K}}_{3;T_1^+,n} ,
\end{align}
where $\wt{\mathbf{K}}_{3;T_1^+,n}$ is an energy-dependent matrix that can be evaluated once. In fact, any parametrization that will have a ${}^3S_1\to{}^3S_1$ contribution to the K matrix will depend on this matrix. Because the spectrum is independent of the $T_1^+$ row chosen, we proceed to omit the row $n$ index in the rest of this work.

Although here we focus our attention on parametrizations that are polynomials in $s$, this pre-computation can be performed for any function of $s$. This is because the $s$ dependence always factorizes out of the $kSm$ space where the subduction must be performed. A scenario where this will not work is parametrizations where one wishes to introduce a pole in $\sigma_k$ in the parametrization of $h_{\ell S}$.

\subsection{Fit results}
\label{app:K3fits}

\begin{table}[t]

\renewcommand{\arraystretch}{1.5}
{ \label{table:ens_info}
\begin{tabular}{c|c|c|c|c|c|c|c|c|c|c}
fit & $\alpha({}^3S_1|{}^3S_1)$ & $\beta({}^3S_1|{}^3S_1)$ & $\gamma({}^3S_1)$ & $\alpha({}^3S_1|{}^3D_1)$& $\alpha({}^3D_1|{}^3D_1)$ & $\alpha({}^3S_1|{}^1P_1)$ & \rm{dof} & Corr& $\chi^2$ & $e^{-(2n_p + \chi^2)/2}$ \\
\hline
1 & $-9(5)$ & & & & & & $11$ & $1$ & $5.1$ & $0.029$ \\
2 & $-8(5)$ & $-0.2(0.4)$ & & & & & $10$ & $\begin{pmatrix}
1.00 & -0.56 \\
& 1.00 \\
\end{pmatrix}$ & $5.0$ & $0.011$ \\
3 & $-7.1(0.8)$ & & $-4(2)$ & & & & $10$ & $\begin{pmatrix}
1.00 & -0.29 \\
& 1.00 \\
\end{pmatrix}$ & $4.7$ & $0.013$ \\
%
4 & $-6(40)$ & $-0.1(0.9)$ & $-4(40)$ & & & & $9$ & $\begin{pmatrix}
1.00 & -0.62 & -0.97 \\
& 1.00 & 0.42 \\
& & 1.00 \\
\end{pmatrix}$ & $4.6$ & $0.005$ \\
5 & $-8(4)$ & $0(1)$ & & $0(2)$ & & & $9$ & $\begin{pmatrix}
1.00 & 0.52 & 0.70 \\
& 1.00 & 0.94 \\
& & 1.00 \\
\end{pmatrix}$ & $4.6$ & $0.005$ \\
6 & $-10(2)$ & &&$0(2)$ & & & $10$ & $\begin{pmatrix}
1.00 & 0.63 \\
& 1.00 \\
\end{pmatrix}$ & $4.7$ & $0.013$ \\
7 & $-8(2)$ & $0(3)$ & & $0(20)$ & $0(5)$ & & $8$ & $\begin{pmatrix}
1.00 & 0.82 & -0.83 & -0.83 \\
& 1.00 & -1.00 & -0.99 \\
& & 1.00 & 1.00 \\
& & & 1.00 \\
\end{pmatrix}$ & $4.4$ & $0.002$ \\
8 & $-10(2)$ & & & $0(1)$ & $0(1)$ & & $9$ & $\begin{pmatrix}
1.00 & 0.02 & -0.53 \\
& 1.00 & -0.29 \\
& & 1.00 \\
\end{pmatrix}$ & $4.7$ & $0.0047$ \\
9 & $-9(2)$ &&& & & $-3(10)$ & $10$ & $\begin{pmatrix}
1.00 & -0.21 \\
& 1.00 \\
\end{pmatrix}$ & $5.0$ & $0.011$ \\
10 & $-8(2)$ & $-0.2(0.3)$ & & & & $1(2)$ & $9$ & $\begin{pmatrix}
1.00 & -0.01 & 0.04 \\
& 1.00 & 0.03 \\
& & 1.00 \\
\end{pmatrix}$ & $5.0$ & $0.0042$

\end{tabular}
}
\caption{ For each fit, we provide the resulting values of the parameters that are left to float in the fit. Those that are fixed to be zero are left blank. In the eighth column, we show the degrees of freedom, defined as ${\rm dof} = 12-n_p$, where $n_{p}$ is the number of parameters in the fit.
}
\label{TABLE:fits}
\end{table}
In this work, we explore a total of 10 parametrizations for the K matrix. For each one, we evaluate the determinant condition, looking for its zeroes in the kinematic region where we have lattice spectra to constrain the fits. As an explicit example, let us define the energy-dependent function,
\begin{align}
{\rm QC}_3 = \det_{I_2 k S m}\left[\left(\mathbf{F}_{3;T_1^+}^{-1}+\mathbf{K}_{3;T_1^+} \right)\times m_\pi^2\right].
\label{eq:QC3_T1}
\end{align}
Note, this is a different function than Eq.~\eqref{eq:QC}.
Given this definition, in Fig.~\ref{fig:QC3T1_v2}, we show the value of this function for a kinematic region where there are three solutions to the quantization condition. This is for the $L/a_s = 16$, namely the smallest volume. In making this plot, we have fixed the two-body parameters to their mean values, as previously mentioned. The $\Kc_3$ parametrization is defined by fixing all parameters to 0 except for $\alpha({}^3S_1|{}^3S_1)$, which is set to $-9$. As we discuss below, this is the preferred value for this parameter when fixing all others to $0$.

As can be seen from Fig.~\ref{fig:QC3T1_v2}, in the range of $E/m_\pi = [4.6,5.2]$, this function has three zeros. This is corresponding to three different finite-volume states, $E_n/m_\pi = 4.63, 4.85,5.14$. From Fig.~\ref{fig:spec}, one sees that these correspond to the first three excited states for the smallest volume. In fact, there is a lower state, which is the ground state, with $E_n/m_\pi = 3.32$.

\begin{figure*}[t]
\begin{center}
\includegraphics[width=.9\textwidth]{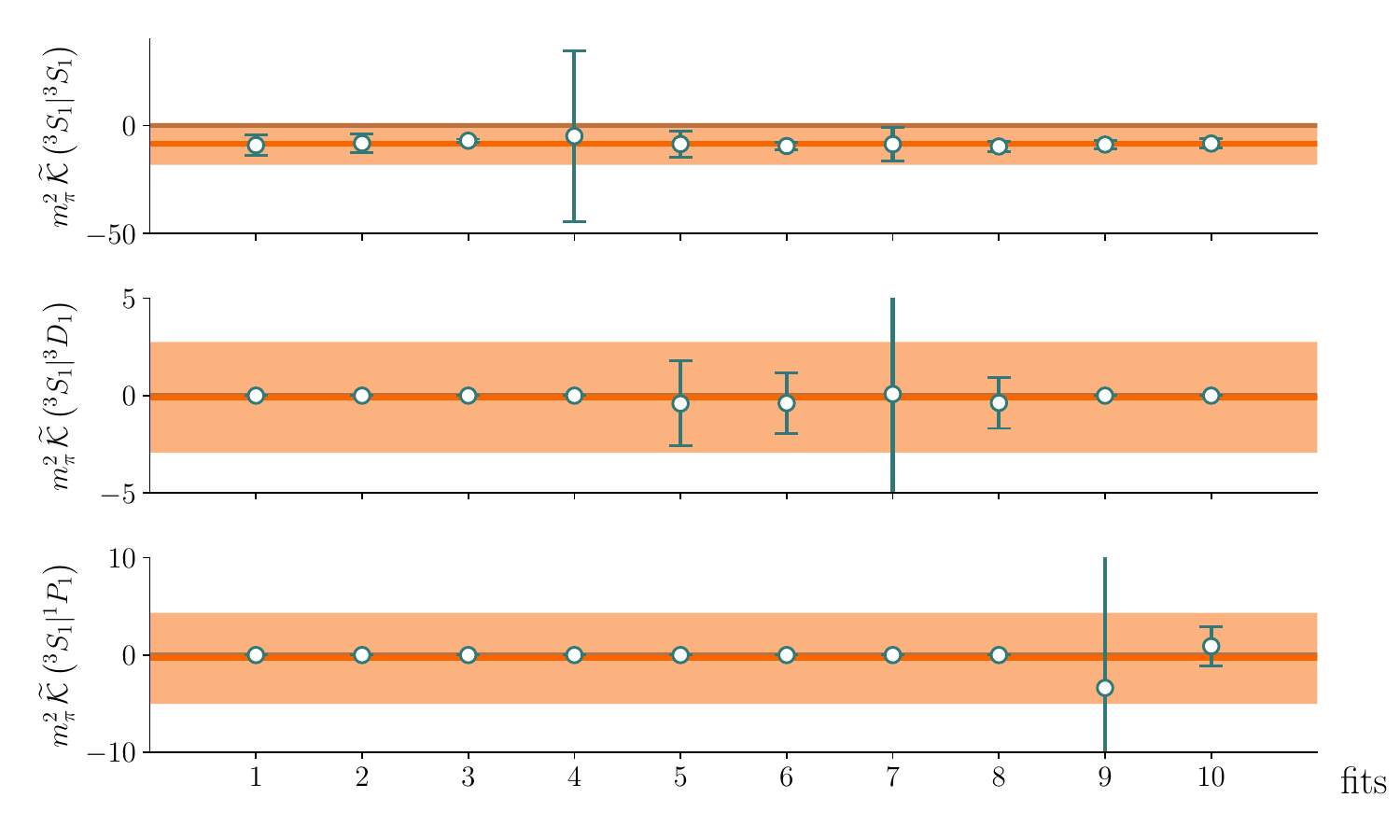}
\caption{ Shown is the values of three components of the reduced $\wt{\Kc}_3$ matrix for a fixed $E=3.4\,m_\pi$. The discrete points are the values obtained from the fits shown in Table~\ref{TABLE:fits}, and the band shows the $1\sigma$ band for the weighted average results.
\label{fig:fits}
}
\end{center}
\end{figure*}
In our fit of this parameter, we fit all 12 lattice-determined energy levels shown in Fig.~\ref{fig:spec}. From this exercise, we get $\alpha({}^3S_1|{}^3S_1)=-9(5),$ with a $\chi^2 =5.1$. We consider 9 other parametrizations which are shown in Table~\ref{TABLE:fits}. For each fit, we give the values of the fitted parameters that are left to float. We also give the degrees of freedom ($\rm{dof}$), the correlation matrix between the parameters, and the relative weight given to each parametrization according to the Akaike information criteria (AIC). The AIC dictates that the weight given to a fit is proportional to $\exp\left[- (-2\,\rm{dof}+\chi^2)/2 \right]$. In this work, ${\rm{dof}} = 12-n_p$, where $n_p$ is the number of parameters for a fit. Since the factor of 12 is a constant for all fits, which modified the normalization of the relative probability, we show $\exp\left[- (2n_p+\chi^2)/2 \right]$ on the last column. From this column, we see that according to the AIC, the first of these fits is the most preferred.

In the left panel of Fig.~\ref{fig:Kmat_amp}, we show in light blue the $1\sigma$ bands for different components of the K matrix coming from all the fits. The opacity of each band is proportional to the probability weight associated with it. In other words, the fainter the band is, the less probable it is. The maroon band is the result is the 1$\sigma$ band for the weight average of all of these fits. The error includes both a statistical as well as systematic error associated with the variation of the model. For this, we follow the expressions presented in Ref.~\cite{Jay:2020jkz,Pefkou:2021fni}. These define the overall uncertainty for a given target observable $A$ as $\sigma_A$ as
\begin{align}
\sigma_A^2 = \sum_f \langle A^2\rangle_f \, p_f -
\left(\sum_f \langle A\rangle_f \, p_f \right)^2 ,
\end{align}
where $\langle A^n\rangle_f$ is the expectation value of $A^n$ for the $f$th fit and $p_f$ is the corresponding fit probability according to the AIC. As an illustration of this, in Fig.~\ref{fig:fits}, we show this procedure applied for the reduced three-body K matrix, $\wt{\Kc}_3$, for a fixed $E=3.4\,m_\pi$. For each component, we show the ten fits from Table~\ref{TABLE:fits}, as well as the AIC weight averaged result.

\section{Amplitude Reconstruction}
\label{app:amp_recon}

Reconstructing the partial wave amplitudes and computing associated observables requires several steps given constrained two- and three-body K matrices. The physical amplitude is written as the sum of two terms $\Dc$ and $\Mc_{3,\df}$, where $\df$ stands for ``divergence free". The first of these two is the so-called \emph{ladder} amplitude, $\Dc$, which encodes all possible pair-wise exchanges, \ie, it does not depend explicitly on $\Kc_3$.\footnote{As discussed in the literature, the integral equation for the ladder amplitude does implicitly depend on the definition of $\Kc_3$ via the choice of the kinematic functions which enter the equations.}. The second term, $\Mc_{3,\df}$, includes all possible insertions of the three-body K matrix, and depends explicitly on the ladder amplitude. Both of these quantities are scheme-dependent as there is freedom as to how one makes this separation~\cite{Hansen:2014eka,Jackura:2022gib}, but the sum of these two is scheme independent.

Following Ref.~\cite{Hansen:2014eka}, it is convenient to write these amplitudes in a basis where one particle is treated as a spectator while the other two form a pair. The full amplitude is obtained by summing over all possible choices of the spectator,
\begin{align}
\mathcal{M}_3
= \sum_{a,b} \Mc_3^{(a,b)} =
\sum_{a,b}
\left[\Dc^{(a,b)}
+\Mc_{3,\df}^{(a,b)}\right],
\label{eq:M3_tot}
\end{align}
where the superscripts $a$ and $b$ parametrize the spectator choice for the initial and final state, respectively. This scheme allows us to define one element contributing to Eq.~\ref{eq:M3_tot}, $\Mc_3^{(a,b)} \equiv \Dc^{(a,b)} + \Mc_{3,\df}^{(a,b)}$. Each of these individual functions are projected to definite $J^P$ quantum numbers, and the resulting $\Dc^{(a,b)J^P}$ and $\Mc_{3,\df}^{(a,b)J^P}$ partial wave amplitudes are defined by a set of coupled linear integral equations which depend on three kinematic variables~\cite{Briceno:2024ehy}.

To determine a single $\Mc_3^{(a,b) J^P}$ partial wave amplitude, we first compute the ladder amplitude from its integral equation. Then, $\Mc_{3,\df}^{(a,b) J^P}$ is determined from its integral equation given the ladder solution and the constrained K matrix as discussed in~\cite{Briceno:2024ehy}. Finally, we follow the procedure outlined in Ref.~\cite{Jackura:2025wbw} to sum over all possible spectators choices and recoupling their results to form the $\Mc_3^{J^P}$ amplitude, from which observables like the intensity distribution in Eq.~\eqref{eq:int} are computed. We summarize the main equations defining these amplitudes and outline these steps in the subsections below, and refer the reader to the references included for more details.

\subsection{The ladder amplitude and its numerical solution}

As shown in Ref.~\cite{Briceno:2024ehy}, the partial wave ladder amplitude, $\Dc^{(a,b)J^P}$, satisfies an integral equation of the form,
\begin{align}
\label{eq:ladder_eq}
\Dc^{(a,b)J^P}(p,k) = \Dc_{0}^{J^P}(p,k) - \Mc_{2}(\sigma_p)\,\cdot\int_{k'}\! \, \Gc ^{J^P}(p,k')\cdot \Dc^{(a,b)J^P}(k',k) \, ,
\end{align}
where $p/k$ are the magnitudes of the spectator momentum of the final/initial state, with the integration defined by
\begin{align}
\int_k &\equiv \int_0^\infty \! \diff k \, \frac{k^2}{(2\pi)^2 \, \omega_k} \, ,
\label{eq:radial_int}
\end{align}
where $\omega_k = \sqrt{m_\pi^2 + k^2}$ is the energy of the integrated spectator. We adopt the notation of Ref.~\cite{Briceno:2024ehy}, where the objects are matrices in $(\ell,S)$-space with the matrix product
\begin{align}
\label{eq:mat_prod}
[A^{J^P}\cdot B^{J^P}]_{\ell 'S',\ell S} &\equiv \sum_{\ell '',S''} A^{J^P}_{\ell 'S',\ell ''S''} B^{J^P}_{\ell ''S'', \ell S} \, .
\end{align}
As previously mentioned, in this work we consider only three wavesets, ${}^{2S+1}\ell_J = \{{}^3S_1,{}^3D_1,{}^1P_1\}$, therefore we have a $3\times 3$ system of linear integral equations. We remind the reader that the isospin of the dipion system is implicit as it is correlated to the $S$ state, that is for isovector dipions we have the waves $(\ell S) = \{(0,1),(2,1)\}$, whereas for isotensor dipions $(\ell S) = \{(1,0)\}$.

The driving term $\Dc_0^{J^P}$ is defined as
\begin{align}
\left[\Dc_{0}^{J^P}(p,k)\right]_{\ell'S',\ell S} &= -\Mc_{2,S'}(\sigma_p) \, \Gc_{\ell'S',\ell S}^{J^P}(p,k) \, \Mc_{2,S}(\sigma_k) \,,
\label{eq:D0}
\end{align}
where $\Mc_{2,S}$ is the partial wave two-body amplitude, defined in terms of its K matrix (see next section), and $\Gc^{J^P}$ is partial wave projection of the one-particle exchange (OPE) propagator~\cite{Jackura:2023qtp},
\begin{align}
\Gc_{I_2'\ell'S',I_2\ell S}^{J^P}(p,k) = J(z_{I_2'}(\sigma_p))\, J(z_{I_2}(\sigma_k)) \left[ \Kc_{\Gc;I_2'\ell'S',I_2\ell S}^{J^P}(p,k) + \Cc_{I_2'\ell'S',I_2\ell S}^{J^P}(p,k)\,Q_0(\zeta_{pk})\right] \, ,
\label{eq:OPE}
\end{align}
where we have explicitly put the isospin dependence due to the cutoff dependence, see Eqs.~\eqref{eq:J_cutoff1} and \eqref{eq:J_cutoff2}. Here $J$ is the cutoff function defined in Eq.~\ref{eq:J_cutoff1}, $Q_0$ is the zero-degree Legendre function of the second kind, $\Kc_{\Gc}^{J^P}$ and $ \Cc^{J^P}$ are derived coefficients, and $\zeta_{pk}$ is a kinematic function which can be written in terms of Lorentz invariant kinematic variables. Since the expressions of these functions is rather lengthy, and they were already tabulated in Ref.~\cite{Jackura:2023qtp}, which also provided a Mathematica notebook for generating these functions, we do not rewrite them here.\footnote{Note is that in Ref.~\cite{Jackura:2023qtp}, the function $\Cc^{J^P}$ was called $\Tc^{J^P}$.}

\begin{figure*}[t]
\begin{center}
\includegraphics[width=0.8\textwidth]{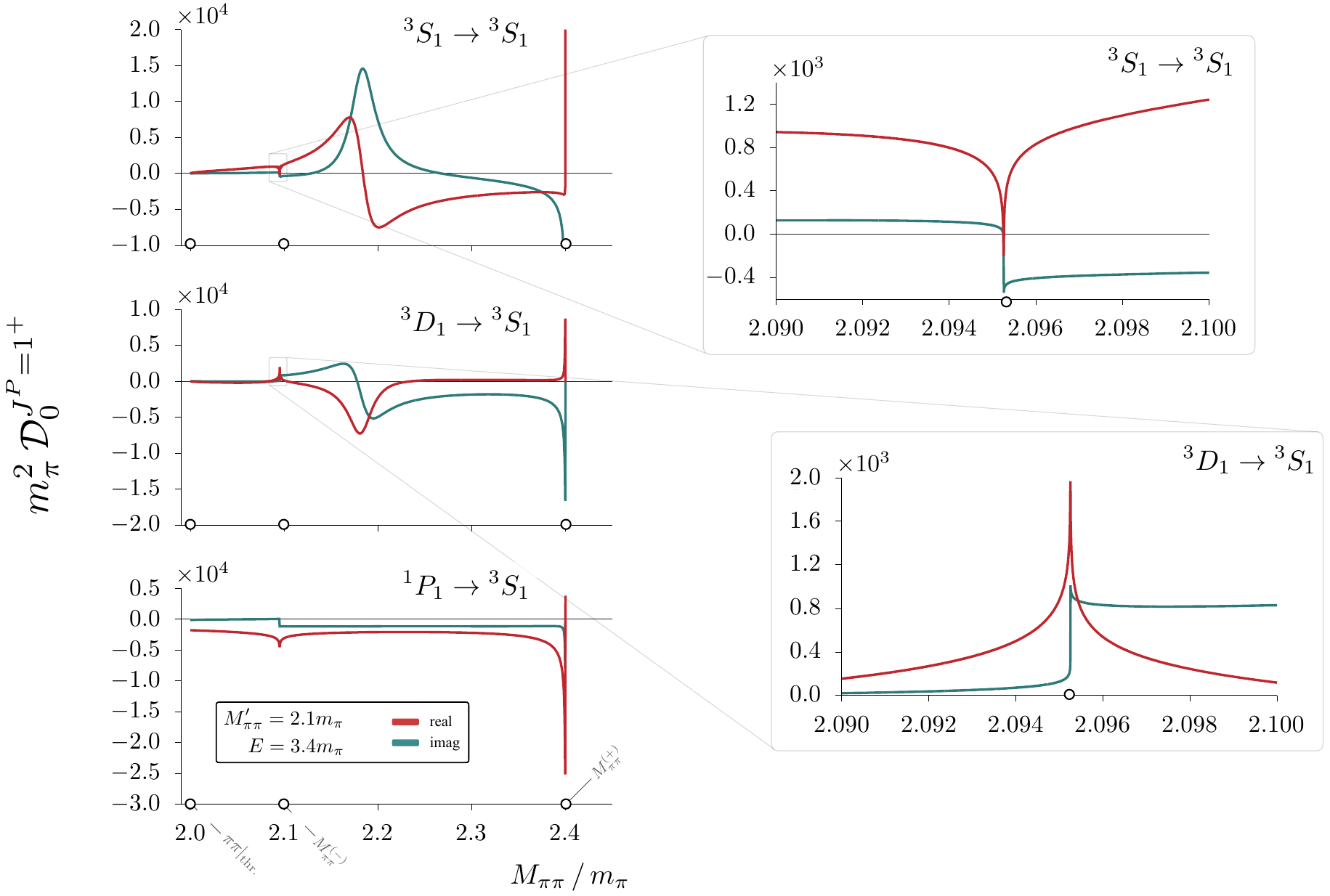}
\caption{Real and imaginary parts for the asymmetric isotensor $J^P=1^+$ $\Dc_0$ amplitudes as a function of $M_{\pi\pi}^2$ for fixed $E=3.4m_\pi$ and $M_{\pi\pi}' = 2.1m_\pi$. Show is the restricted waveset used in this analysis, ${}^{2S+1}\ell_J = \{{}^3S_1,{}^3D_1,{}^1P_1\}$. The insets on the ${}^3S_1\to{}^3S_1$ and ${}^3D_1\to{}^3S_1$ amplitudes highlight the logarithmic singularity due to the OPE mechanism.
\label{fig:D0_plots}
}
\end{center}
\end{figure*}
We show plots of the driving term $\Dc_0$ in Fig.~\ref{fig:D0_plots} for the scattering channel of interest in this work, the isotensor $J^P = 1^+$ system, with the relevant waves used in the analysis. Plots are shown as a function of $M_{\pi\pi} \equiv \sqrt{\sigma_{k}}$ for the same kinematics shown in Fig.~\ref{fig:Kmat_amp}(b), that is for fixed $E = 3.4m_\pi$ and the final state $M_{\pi\pi}' \equiv \sqrt{\sigma_{k'}} = 2.1m_\pi$. The physical $M_{\pi\pi}$ region is defined by $2m_\pi \le M_{\pi\pi} \le E - m_\pi$, with the maximum for the case shown being $E - m_\pi = 2.4m_\pi$. The OPE gives singularities at
\begin{align}
M_{\pi\pi}^{(\pm)} = \frac{1}{\sqrt{2}} \left[\,3m_\pi^2 + E^2 - M_{\pi\pi}'^2 \pm \lambda^{1/2}(E^2,m_\pi^2,M_{\pi\pi}'^2)\sqrt{1-\frac{4m_\pi^2}{M_{\pi\pi}'^2}} \, \right]^{1/2} \, ,
\end{align}
which for our kinematics is $M_{\pi\pi}^{(-)} \approx 2.09526 m_\pi$ and $M_{\pi\pi}'^{(+)} \approx 2.39998 m_\pi$. Notice that the $M_{\pi\pi}^{(-)}$ singularity is softened for $S=1$ dipion pairs, while the $M_{\pi\pi}^{(+)}$ singularity is enhanced due to the divergence at the physical boundary $M_{\pi\pi} \sim E-m_\pi$. See Refs.~\cite{Jackura:2018xnx,Jackura:2023qtp,Jackura:2025wbw} for more details.

Equation~\eqref{eq:ladder_eq} is a linear system of Fredholm integral equations of the second kind, where the kernel is non-degenerate due to the non-analytic behavior of the OPE. We simplify Eq.~\eqref{eq:ladder_eq} by defining an intermediary function $d^{J^P}$ as
\begin{align}
\Dc_{\ell'S',\ell S}^{J^P}(p,k) &\equiv -\Mc_{2,S'}(\sigma_p) \, d_{\ell'S',\ell S}^{J^P}(p,k) \, \Mc_{2,S}(\sigma_k) \, ,
\end{align}
to remove the explicit two-body scattering dependency from both the initial and final state. We obtain numerical solutions of the resulting integral equation for $d^{J^P}$ by following the Nystr\"om method, where the integral is approximated by a Gauss-Legendre quadrature rule of order $N_k$,
\begin{align}
\label{eq:d_approx}
d^{J^{P}}(p,k) = -\Gc^{J^{P}}(p,k) - \sum_{j=0}^{N_k-1} \Qc^{J^{P}}(p,k_j) \cdot d^{J^{P}}(k_j,k) \, ,
\end{align}
where $k_j$ are the Gauss-Legendre nodes in momentum space, $\Qc^{J^P}$ is the kernel
\begin{align}
\Qc^{J^P}(p,k_j) \equiv
\Gc^{J^P}(p, k_j) \cdot \Mc_2\left(\sigma_{k}(k_j)\right)
\, \frac{ k_{j}^2}{(2\pi)^2 \,\omega_{k_{j}}}\, \Delta_{j},
\end{align}
with $\Delta_j$ being the weights from the quadrature rule. We evaluate $p$ and $k$ on the momentum partition $\{k_j\}_{j=0}^{N_k-1}$, converting Eq.~\eqref{eq:d_approx} into a square linear algebraic system of order $N_c\times N_k$ in the combined channel and momentum space. We numerically solve the system, choosing sufficiently large $N_k$ to ensure convergence of the solution, see Ref.~\cite{Briceno:2024ehy, Dawid:2023jrj} for more details.

As can be seen from Fig.~\ref{fig:D0_plots}, the OPE singularity for this system is only resolvable in a small kinematic window. The momentum resolution of the resulting amplitudes is inversely proportional to the number of mesh points ($N_k$) used in solving the integral equations. Even though we see good convergence in our numerical solutions to the integral equations for the majority of the kinematic region shown in Fig.~\ref{fig:Kmat_amp}, the OPE cusp is just barely resolvable for the matrix sizes reached in this analysis.

\subsection{The divergence-free three-body amplitude}
\label{app:int_eq}

Given the numerical solution for $\Dc^{J^P}$ and the constrained three-body K matrix, we can reconstruct the remaining divergence-free amplitude from its system of integral equations. For separable K matrices of the form of Eq.~\eqref{eq:K3.separable}, one can write $\Mc_{3,\df}$ the integral equations are algebraically solvable in terms of integrals over known functions. Following Ref.~\cite{Briceno:2024ehy}, $\Mc_{3,\df}^{(a,b) J^P}$ can be computed by
\begin{align}
\Mc_{3,\df}^{(a,b)J^P}(p,k) = \Lc^{\prime\,J^P}(p) \cdot \Tc^{J^P} \cdot \Lc^{J^P}(k) \,,
\label{eq:M3_df}
\end{align}
where $\Lc$ and $\Lc^{\prime}$ are initial and final state rescattering functions, respectvely,
\begin{align}
\Lc^{\prime\,J^P}(p) & = h(p) - \int_k \, \Mc_2 (\sigma_{p}) \cdot \Gamma^{J^P}(p,k) \cdot h(k)
- \int_k \int_{k'} \Dc^{(a,b)J^P} (p,k') \cdot \Gamma^{J^P}(k',k) \cdot h(k) \, , \\[5pt]
\Lc^{J^P}(k) & = h(k) - \int_p \, h(p) \cdot \Gamma^{J^P}(p,k) \cdot \Mc_2 (\sigma_{k})
- \int_p \int_{p'} h(p) \cdot \Gamma^{J^P}(p,p') \cdot \Dc^{(a,b)J^P} (p',k) \, ,
\end{align}
which dress the function $\Tc$ which is given in terms of $\wt{\Kc}_3$,
\begin{align}
\Tc^{J^P}(s) &= \frac{1}{1 + \wt{\Kc}_3^{J^P}(s) \cdot \Fc^{J^P}(s) } \cdot \wt{\Kc}_3^{J^P}(s) \, .
\end{align}
The functions $\Gamma$ and $\Fc$ are defined by
\begin{align}
\Gamma^{J^P}(p,k) &= \frac{(2\pi)^2\,\omega_k}{k^2} \, \delta(p - k) \, \wt{\rho}(\sigma_p) + \Gc^{J^P}(p,k) \, ,
\\
\label{eq:F}
\Fc^{J^P}(p,k) &= \, \wt{\rho}(\sigma_p) \, \Lc^{\prime\,J^P}(p,k) + \int_{k'} \, \Gc(p,k') \cdot \Lc^{\prime\,J^P}(k',k) \, ,
\end{align}
which can be regarded as containing all rescatterings in the intermediate state.

As discussed in Sec.~\ref{app:spec_anal}, the presence of poles in the Breit-Wigner parametrization requires a modification of the definition of the phase space in the integral equations. Making the channel indices explicit, $\wt{\rho}$ can be written as
\begin{align}
[\wt{\rho}]_{I_2'\ell 'S',I_2\ell S}
& =
J(z_{I_2}(\sigma_k))\left[-i\, \rho(\sigma_k)
+
\delta_{I_2,1}
\Ic_{\rm PV}\frac{ m_\pi^2 }{32 \pi \, q_k^{\star 2} }
\right]
\,\delta_{I_2',I_2}
\,\delta_{\ell ',\ell }
\,\delta_{S', S} \, ,
\label{eq:rhoshift}
\end{align}
with $\rho(\sigma_k) = q_k^\star / (16\pi\sqrt{\sigma_k})$.

\subsection{Symmetrizing partial wave amplitudes}
\label{app:symm}

\begin{figure*}[t]
\begin{center}
\includegraphics[width=\textwidth]{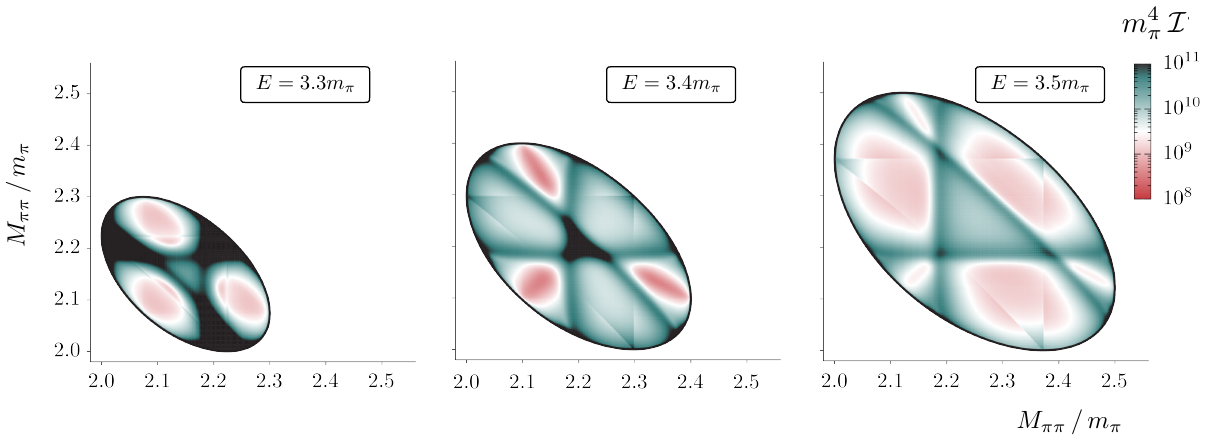}
\caption{Dalitz plots of the intensity distributions in Eq~\eqref{eq:int} for the $J^P=1^+$ driving term, Eq.~\eqref{eq:D0}, for $E/m_\pi = \{3.3,3.4,3.5\}$. Initial and
final state kinematics are fixed to be the same.\label{fig:OPE_symm}
}
\end{center}
\end{figure*}

Finally, we follow the procedure outlined in Ref.~\cite{Jackura:2025wbw} to symmetrize the resulting asymmetric amplitudes to obtain a definite $J^P$ partial wave amplitude which is independent of the spectator choice. The symmetric partial wave amplitude is given by the sum
\begin{align}
\Mc_{3,I_2'\lambda',I_2\lambda}= \sum_{a,b}{\Rc}_{I_2'\lambda'}^{(a)} \cdot {\Mc}_{3}^{(a,b)} \cdot {\Rc}_{I_2 \lambda}^{(b)\,\top}
\label{eq:symm}
\end{align}
where $a/b$ sum over the three possible permutations of the spectator choice for the initial/final state. The $I_3(J^P)$ recoupling coefficients $\Rc$ are matrices which coupled the set $\{I_2,\ell,S\}$ defined with respect to the spectator $u$ to some desired quantum numbers $I_2\lambda$, with $\lambda$ an effective helicity label for a dipion pair. One must still choose a particle to define these quantum numbers. Without loss of generality, we choose the first pion. The recoupling coefficients are defined as a product of pure isospin and angular momentum coefficients,
\begin{align}
\Rc_{I_2 \lambda; I_2^{(a)} \ell^{(a)} S^{(a)}}^{(a)I_3(J^P)} = \Rc_{I_2; I_2^{(a)}}^{I_3} \, \Rc_{\lambda; \ell^{(a)} S^{(a)}}^{J^P} \, ,
\end{align}
where the superscripts on the indices indicate which respective spectator the quantum number is defined. The isotensor recoupling coefficients for three pions are given by the identity if the spectator and target pion are the same, and
\begin{align}
\Rc_{I_2 I_2^{(a)}}^{I_3 = 2} \, (\mathrm{cyclic}) = \frac{1}{2}\begin{pmatrix}
-1 & \sqrt{3} \\
-\sqrt{3} & -1
\end{pmatrix} \, , \qquad
\Rc_{I_2 I_2^{(a)}}^{I_3 = 2} \, (\mathrm{anticyclic}) = \frac{1}{2}\begin{pmatrix}
-1 & -\sqrt{3} \\
\sqrt{3} & -1
\end{pmatrix} \, ,
\end{align}
where the cyclic coefficient is for when the $u$ spectator and target pion are off by a cyclic permutation, and similarly, the anticyclic coefficient is for when they are off by an anticyclic permutation. The $J^P = 1^+$ recoupling coefficent is
\begin{align}
\Rc^{J^P = 1^+}_{\lambda; \ell^{(a)} S^{(a)}} = \sqrt{2S^{(a)}+1} \sum_{\mu = -J}^{J} d_{\lambda\mu}^{(J)}(\Omega) \, d_{\mu 0}^{(S^{(a)})}(\vartheta) \, \Pc_\mu({}^{2S^{(a)}+1}\ell^{(a)}_{J})
\end{align}
where $d$ are the Wigner little d matrices and $\Pc_{\mu}({}^1P_1) = \delta_{\mu 0}$ while the triplet state is
\begin{align}
\Pc_{\mu}({}^3\ell_1) = \sqrt{\frac{1}{3}}\,\begin{cases}
\, 1 \, , \qquad &\textrm{for }\ell = 0 \, , \\[5pt]
\, \sqrt{\frac{1}{2}} \lvert\mu\rvert - \sqrt{2} \delta_{\mu 0} \, , \qquad &\textrm{for } \ell = 2 \, .
\end{cases}
\end{align}
The arguments $\Omega$ and $\vartheta$ are functions of kinematics which depend on if the target pion is identical, a cyclic permutation, or anticyclic permutation to the spectator pion $u$ (see Ref.~\cite{Jackura:2025wbw} for details and their functional form).

As an illustration of this procedure, in Fig.~\ref{fig:OPE_symm} we show the intensity using the driving term of the integral equations, Eq.~\eqref{eq:D0}, as the input. Dalitz plots are shown for $E/m_\pi = 3.3, 3.4,3.5$, with the initial state invariant masses mirrored to the final state. The OPE singularities illustrated in Fig.~\ref{fig:D0_plots} along with their reflections are noticeable in these plots, However notice that their regions are quickly damped out by their corresponding two-pion partial wave amplitudes.

\end{document}